\documentclass[twoside,singlecolumn,floatfix]{revtex4}

\usepackage{epsfig}
\usepackage{amsmath}
\usepackage{amssymb}
\usepackage{bm}
\usepackage{times}
\usepackage{color} 

\def\picdirectory{.}

\definecolor{dark-blue}{rgb}{0,0,0}
\definecolor{light-blue}{rgb}{0,0,0} 
\definecolor{light-green}{rgb}{0,0,0} 
\definecolor{light-brown}{rgb}{0,0,0}
\definecolor{light-grey}{rgb}{0,0,0} 
\definecolor{dark-yellow}{rgb}{0,0,0} 
\definecolor{light-yellow}{rgb}{0,0,0} 
\definecolor{lila}{rgb}{0,0,0}
\definecolor{yellow}{rgb}{0,0,0}
\definecolor{red}{rgb}{0,0,0}
\definecolor{green}{rgb}{0,0,0}
\definecolor{blue}{rgb}{0,0,0}
\definecolor{dark-green}{rgb}{0,0,0}
\definecolor{blue-green}{rgb}{0,0,0}
\definecolor{red-green}{rgb}{0,0,0}
\definecolor{black}{rgb}{0,0,0}

\usepackage{citesort}

\newcommand{\avg}[1]{\langle{#1}\rangle}
\newcommand{\req}[1]{(\ref{#1})}
\newcommand{\beq}{\begin{equation}}
\newcommand{\eeq}{\end{equation}}
\newcommand{\beqas}{\begin{eqnarray*}}
\newcommand{\eeqas}{\end{eqnarray*}}
\renewcommand{\vec}[1]{\mbox{\boldmath$#1$}}
\newcommand{\lab}[1]{\label{#1}}

\begin{document}

\title{A driven two-dimensional granular gas with Coulomb friction}

\author{Olaf Herbst (1), 
        Raffaele Cafiero (2,3), 
        Annette Zippelius (1),
        Hans J\"urgen Herrmann (2,3),
        Stefan Luding (3,4)
}
       
\affiliation{
(1) Institut f\"ur Theoretische Physik,
    Friedrich-Hund-Platz 1, 37077 G\"ottingen, GERMANY \\
    e-mail: {\tt herbst@theorie.physik.uni-goettingen.de} \\
(2) P.M.M.H., Ecole Sup\`erieure de Physique et de Chimie
    Industrielles (ESPCI),\\
    10, rue Vauquelin-75251 Paris cedex 05, FRANCE \\
(3) Institute for Computer Physics,
    Pfaffenwaldring 27, 70569 Stuttgart, GERMANY \\
(4) Particle Technology, DelftChemTech, TUDelft,
    Julianalaan 136, 2628 BL Delft, NETHERLANDS \\
    e-mail: {\tt s.luding@tnw.tudelft.nl}
}

\date{\today}

\begin{abstract}
  We study a homogeneously driven granular gas of inelastic hard
  particles with rough surfaces subject to Coulomb friction.  The
  stationary state as well as the full dynamic evolution of the
  translational and rotational granular temperatures are investigated
  as a function of the three parameters of the friction model. Four
  levels of approximation to the (velocity-dependent) tangential
  restitution are introduced and used to calculate translational and
  rotational temperatures in a mean field theory.  When comparing
  these theoretical results to numerical simulations of a randomly
  driven mono-layer of particles subject to Coulomb friction, we find
  that already the simplest model leads to qualitative agreement, but
  only the full Coulomb friction model is able to reproduce/predict
  the simulation results quantitatively for all magnitudes of
  friction. In addition, the theory predicts two relaxation times for
  the decay to the stationary state.  One of them corresponds to the
  equilibration between the translational and rotational degrees of
  freedom.  The other one, which is slower in most cases, is the
  inverse of the common relaxation rate of translational and
  rotational temperatures.
\end{abstract}


\maketitle

\section{Introduction}\label{sec:introduction}
Granular media are collections of macroscopic particles with arbitrary
shape, rough surfaces, and dissipative interactions
\cite{herrmann98,vermeer01,kishino01,poschel03}.  Many phenomenona are
well reproduced by model granular media, where spheres are used
instead of other, possibly more realistic shapes.  In order to study
such model systems, kinetic theories
\cite{jenkins85b,lun87,lun91,goldshtein95,jenkins97b,huthmann97,noije98d,noije99,herbst00,jenkins02,luding03c,poschel03,goldhirsch04}
and numerical simulations
\cite{walton86,campbell90,louge94,luding95b,luding98d,mcnamara98,luding03c,poschel03,moon03b}
have been applied for special boundary conditions and a variety of
interesting experiments have been performed, see for example
\cite{olafsen98,olafsen99,losert99,baxter03,geminard04}.  The dynamics
of the system is usually assumed to be dominated by instantaneous
two-particle collisions. These collisions are dissipative and
frictional, and conserve linear and angular momentum while energy is
not conserved.  In the simplest model, one describes inelastic
collisions by a normal restitution coefficient $r$ only. However,
surface roughness and friction are important
\cite{luding95b,huthmann97,luding98d,mcnamara98,herbst00,cafiero00},
since they allow for an exchange of translational and rotational
energy and influence the overall dissipation.  In the standard
approach \cite{jenkins85b,mcnamara98,huthmann97}, surface roughness is
accounted for by a constant tangential restitution coefficient 
$r_t$, which is defined in analogy to $r$ in the tangential direction.
A more realistic friction law involves the Coulomb friction
coefficient $\mu$ \cite{walton86,foerster94,luding95,luding98c}, so
that the tangential restitution $r_t(\gamma)$ depends on the impact
angle $\gamma$, i.e.\ the angle between the contact normal and the
relative velocity of the contact points.

Recently, Jenkins and Zhang \cite{jenkins02} proposed a kinetic theory
for frictional, nearly elastic spheres in the limit of small friction
coefficient $\mu$.  They introduced an effective coefficient of normal
restitution by approximately relating the rotational temperature to
the translational one. Thereby the kinetic theory for slightly
frictional, nearly elastic spheres has the same structure as that for
frictionless spheres.  Also for small $\mu$, Goldhirsch et
al.\,\cite{goldhirsch04} showed that an infinite number of
spin-dependent densities is needed to describe the dynamics of
frictional spheres and that the distribution of rotational velocities
is non-Gaussian.  A mean field theory for three dimensional cooling
systems of rough particles with Coulomb friction was proposed in
\cite{herbst00} and found to be in very good agreement with computer
simulations for a wide range of parameters. A systematic theoretical
study of driven systems over the whole range of dissipation and
friction parameters is not available to our knowledge.

In the following, we propose a mean-field (MF) theory of homogeneously
driven rough particles that accounts for Coulomb friction (i.e.\ a
non-constant $r_t$) on different levels of refinement.  The most
accurate description parallels the three-dimensional (3D) results
\cite{herbst00} for freely cooling systems. In addition, we present
different levels of approximation to the full model and discuss their
shortcomings in MF theory. The homogeneous driving used here is the
same as in other recent studies of driven systems
\cite{cafiero00,luding03c}.

To test our analytical results we have performed numerical simulations
of a randomly driven mono-layer of spheres, using an Event Driven (ED)
algorithm \cite{lubachevsky91,luding98d,mcnamara98,cafiero00}.  One
key result is that, via $r_t(\gamma)$, all parameters of the collision
model affect the evolution of the translational and rotational degrees
of freedom (temperatures) of the system. Only the full MF theory is
able to quantitatively predict the system behavior for the whole
parameter range.

The model system is introduced in section \ref{sec:model}. The
distribution of impact angles, as affected by translational and
rotational degrees of freedom, is computed in section
\ref{sec:impact.angle}. The standard approach with constant tangential
restitution is briefly reviewed, before we introduce three levels of
approximation and the full MF theory in section
\ref{sec:DGL.Mean.Field}. In section \ref{sec:steadystate} we discuss
the stationary state and and in section
\ref{sec:approach.to.steady.state} the dynamic evolution towards the
stationary state. In both sections we compare the predictions of full
MF theory and its approximations to simulations.  Finally we present a
summary and conclusions in section \ref{sec:conclusions}.

\section{Model}\label{sec:model}

\begin{figure}[htb]
\begin{center}
\hspace{-.4cm}~\epsfig{file=\picdirectory/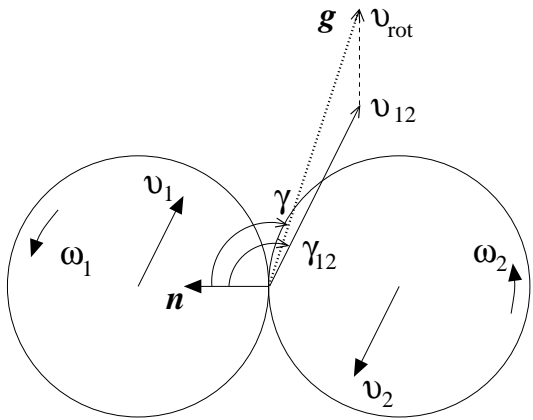,height=5.5cm,angle=0}
\end{center}
\caption{Schematic drawing of two-particle contact in the center of
  mass reference frame.  Shown are the relative velocity
  $\boldsymbol{g}$ of the contact points, the impact angle $\gamma$ of
  the contact points, and the angle $\gamma_{12}$ between the relative
  translational velocity of the particles and their contact normal.}
\label{fig:2pcoll}
\end{figure}

The model system contains $N$ three-dimensional spheres of diameter
$2a$, mass $m$, and moment of inertia $I$ interacting via a hard-core
potential. The particles are confined to a two-dimensional (2D) square
with periodic boundary conditions.  The linear box size is $L$ and the
area (volume) $V=L^2$.  The moment of inertia can be expressed using
the shape factor
\begin{equation}\lab{eq:q}
q:=\frac{I}{ma^2}\,.
\end{equation}
For spheres with a homogeneous mass distribution $q=2/5$.
Inelasticity and roughness are described by a coefficient of normal
restitution $r$, the Coulomb friction law with coefficient of friction
$\mu$, and a coefficient of tangential restitution $r_t$ which depends
on $r$, $\mu$, and the impact angle $\gamma$ for sliding contacts, or
on a maximum tangential restitution $r_t^m$ for sticking contacts,
when some ``tangential elasticity'' becomes important.  In a collision
of two particles $i=1$ and $2$ with positions $\vec{r}_i$, contact
normal $\vec{n}=(\vec{r}_1-\vec{r}_2)/(2a)$, angular velocities
$\vec{\omega}_i$ and relative translational velocity
$\vec{v}_{12}=\vec{v}_1-\vec{v}_2$ (see Fig.\ \ref{fig:2pcoll}), their
velocities after the collision are related to the velocities before
the collision, through a collision matrix
\cite{walton93,luding95b,luding98c} which is derived from the
conservation laws for linear and angular momentum, energy/dissipation
balance, and Coulomb's law.  This three parameter model is able to
reproduce the experimental measurements on colliding spheres of
various materials \cite{foerster94,labous97}.

\subsection{Collision rules}\label{sec:collision.rules}

The collision rules are most transparent when written in terms of the
relative velocity of the contact point in the center-of-mass reference
frame
\begin{equation}\label{eq:def.g}
\vec{g}_{} =  \vec{v}_1-\vec{v}_2- 
 a\left(\vec{\omega}_1+\vec{\omega}_2 \right) \times \vec{n}\,.
\end{equation}
We decompose $\vec{g}=\vec{g}_n+\vec{g}_t$ into its normal and
tangential components with respect to $\vec{n}$, $\vec{g}_n=
(\vec{g}\cdot\vec{n}) \vec{n}$ and $\vec{g}_{t}= \vec{g}-\vec{g}_n$.
The change of normal momentum of particle 1, denoted by $\Delta
\vec{P}^{(n)}$ is the same as for smooth particles
\begin{equation}\lab{eq:linearmomentumchange}
\Delta
\vec{P}^{(n)}=-(m/2)(1+r)\vec{g}_{n}\,. 
\end{equation}

 The change of tangential momentum
\begin{equation}\lab{eq:tangentialmomentumchange}
\Delta \vec{P}^{(t)}=-\frac{q}{q+1}m(1+r_t)\vec{g}_{t}
\end{equation}
is, in general, a function of the impact angle $\gamma$.  Coulomb
friction can be expressed \cite{walton93} in terms of a coefficient of
tangential restitution
\begin{equation} \label{eq:betagamma}
r_t(\gamma) = \min \left [ r_t^C(\gamma), r_t^m \right ]~,
\end{equation}
which is a function of the impact angle $\gamma$ between $\vec{g}_{}$
and $\vec{n}$. Here $r_t^m$ is the coefficient of maximum tangential
restitution, with $-1\leq \! r_t^m \! \leq \! 1$ to ensure that energy
is not created.  The quantity $r_t^C(\gamma)$ is determined using
Coulomb's law
\begin{equation}
r_t^C(\gamma) = - 1 - \frac{q+1}{q} \mu (1+r) \cot \gamma ~,
\label{eq:betagamma.nurCoulomb}
\end{equation}
with the impact angle $\pi/2 < \gamma \le \pi$ so that $\cos \gamma =
\vec{g}\cdot\vec{n}/|\vec{g}|$ is always negative
\cite{foerster94,luding95b,luding98c}. Here, we have simplified the
tangential contacts in the sense that exclusively either Coulomb
friction applies, i.e.\ $\Delta P^{(t)} = \mu \Delta P^{(n)}$, or
constant tangential restitution with the maximum tangential
restitution coefficient $r_t^m$.  Coulomb friction is effective when
the relative tangential velocity is large, whereas tangential
restitution applies for low tangential velocities.

\begin{figure}[htb]
\begin{center}
\vspace{0.8cm}~\\
\epsfig{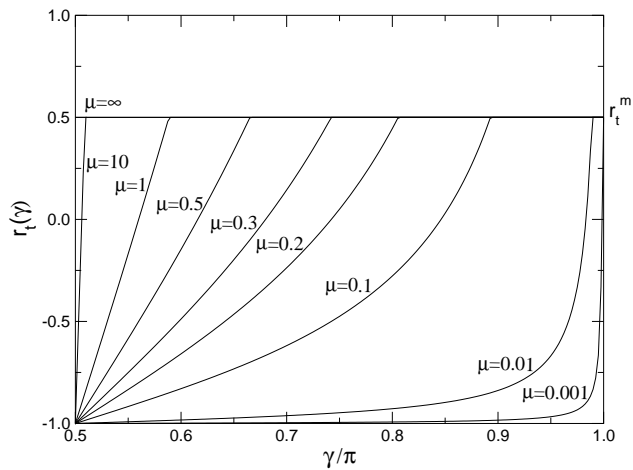}
\end{center}
\caption{Tangential restitution $r_t$ as function of the
impact angle $\gamma$ for different values of the coefficient
of friction $\mu$.
   }
\label{fig:r_t}\label{fig:BetaVonGamma}
\end{figure}

Note that in the general case $\vec{v}_{\rm rot}=
-a(\vec{\omega}_1+\vec{\omega}_2) \times \vec{n}\neq 0$, so that the
{} angle $\gamma_{12}$ between the contact normal $\vec{n}$ and the
relative translational velocity $\vec{v}_{12}=\vec{v}_{1}-\vec{v}_{2}$
is different from the impact angle $\gamma$ of the contact points, see
Figs.\ \ref{fig:2pcoll} and \ref{fig:r_t}.  In the following we will
refer to $\gamma$ when we talk about the impact angle.

\subsection{Driving model}\label{sec:driving.model}

The driving of a granular material can be realized by moving walls,
see Ref.\ \cite{herrmann98} and references therein, corresponding to a
local heating \cite{luding99b,soto99,soto01}, or the system can
alternatively be driven by a global homogeneous, random energy source
in different variations
\cite{puglisi98,bizon99,noije98c,noije98d,noije99,cafiero00,cafiero02}.
We choose homogeneous translational driving here and modify the
velocity of particle $i$ at each time of agitation $t$ such that
\begin{eqnarray}\label{eq:hom.driving}
\boldsymbol{v}'_{i}(t) & = & \boldsymbol{v}_i(t) + v_{\rm dr} ~\boldsymbol{\xi}_i(t)
\end{eqnarray}
where the prime on the left hand side indicates the value after the
driving event.  Measuring masses in units of the particle mass $m$,
the driving velocity $v_{\rm dr}$ sets the time (velocity) scale and
defines the driving temperature $T_{\rm dr}:=m v_{\rm dr}^2$.  The
components of the vector $\boldsymbol{\xi}_i(t)$, $\xi_{i,x}(t)$ and
$\xi_{i,y}$(t), are uncorrelated Gaussian random numbers with zero
mean and variance
\begin{equation}
 \avg{\xi_{i,k}(t)\,\xi_{j,l}(t')}_{\{\xi\}}=\delta_{ij}\delta_{kl}\delta(t-t')~,
\end{equation}
where $\delta_{ij}$ and $\delta_{kl}$ are Kronecker deltas and
$\delta(t-t')$ is the Dirac delta function.  The stochastic driving
rule in Eq.\ (\ref{eq:hom.driving}) leads to an average rate of change
of temperature
\begin{equation}
\Delta T/\Delta t = H_{\rm dr} ~,
{\rm ~~ with  ~~} H_{\rm dr} = f_{\rm dr} T_{\rm dr} ~,
\label{eq:Hdr}
\end{equation}
after every driving time-step $\Delta t = f_{\rm dr}^{-1}$.

\subsection{Simulations}

We have performed simulations of a randomly driven mono-layer of
spheres, using an Event Driven (ED) algorithm
\cite{lubachevsky91b,luding95b,luding98d,cafiero00}, and compared the
results with the MF predictions, see also Refs.\ 
\cite{noije98c,noije98d,herbst00,cafiero00,cafiero02}.  Every
simulation is equilibrated without driving with $r=1$ and in the
smooth surface limit $r_t^m=-1$.  Then inelasticity, friction and
driving are switched on, according to the rules defined above. The
problem of the inelastic collapse characteristic of the ED algorithm
\cite{mazighi94,mcnamara94}, is handled by using normal restitution
coefficients dependent on the time elapsed since the last event
\cite{luding96e,luding98f,luding03}.  The frequency of driving is
chosen such that it is larger than or comparable to the typical
collision frequency per particle, both initially and in steady state.
Varying the driving frequency to much larger values did not affect the
simulation results, whereas the use of a much smaller driving rate
caused different results due to the slow input of energy.

\section{Impact-angle probability distribution}
\label{sec:impact.angle}
\label{sec:pgamma}

In the following we shall discuss various levels of approximation to
the collision rules given in Eqs.\ (\ref{eq:betagamma}) and
(\ref{eq:betagamma.nurCoulomb}).  One possibility to simplify the
collision rules is to consider tangential restitution averaged over
all impact angles $\gamma$, thereby reducing the problem to one with a
constant coefficient of tangential restitution. For that purpose we
need to know the probability distribution of impact angles.

\begin{figure}[htb]
\begin{center}
  \epsfig{file=\picdirectory/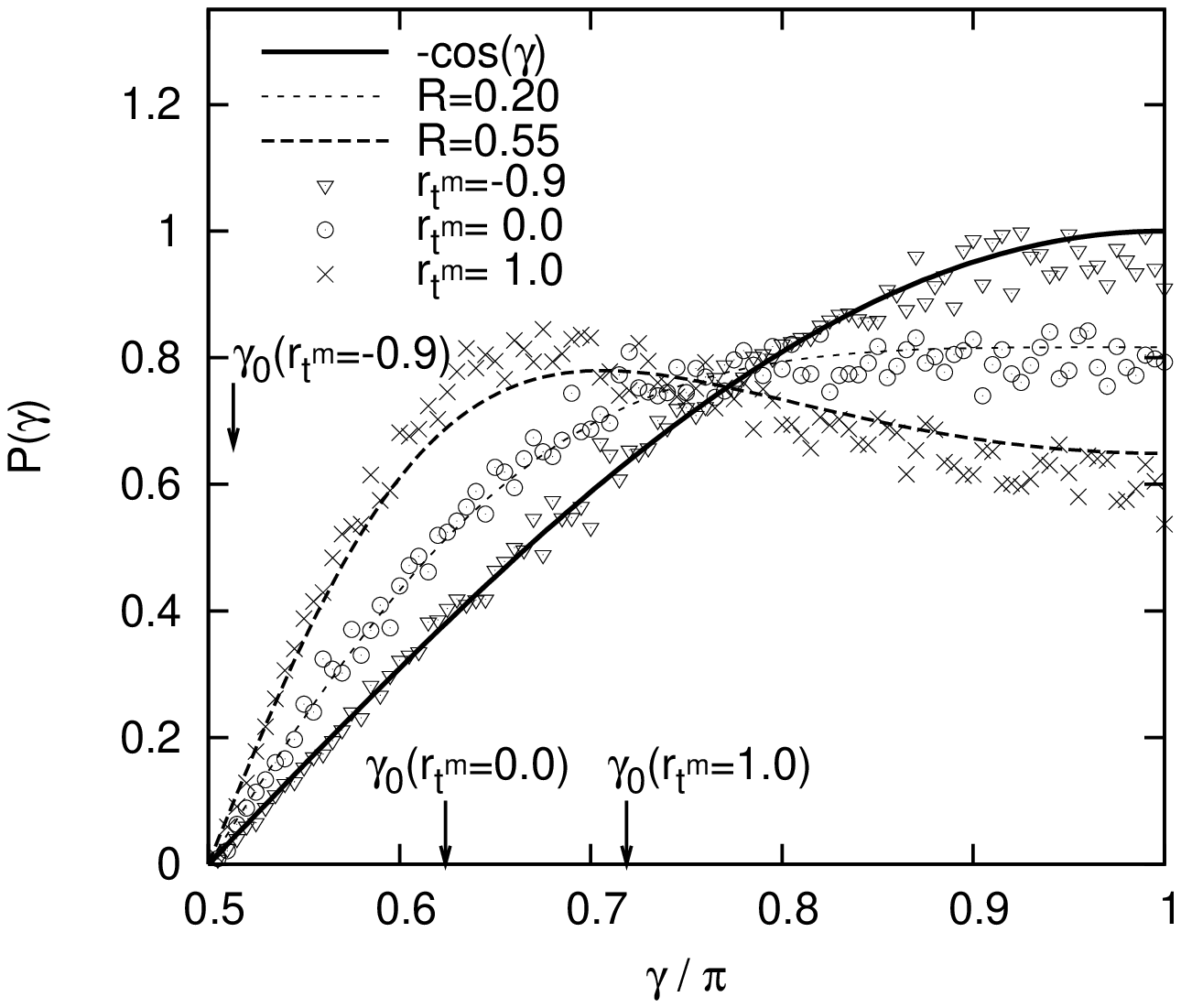,height=7.5cm,angle=-90}
  \epsfig{file=\picdirectory/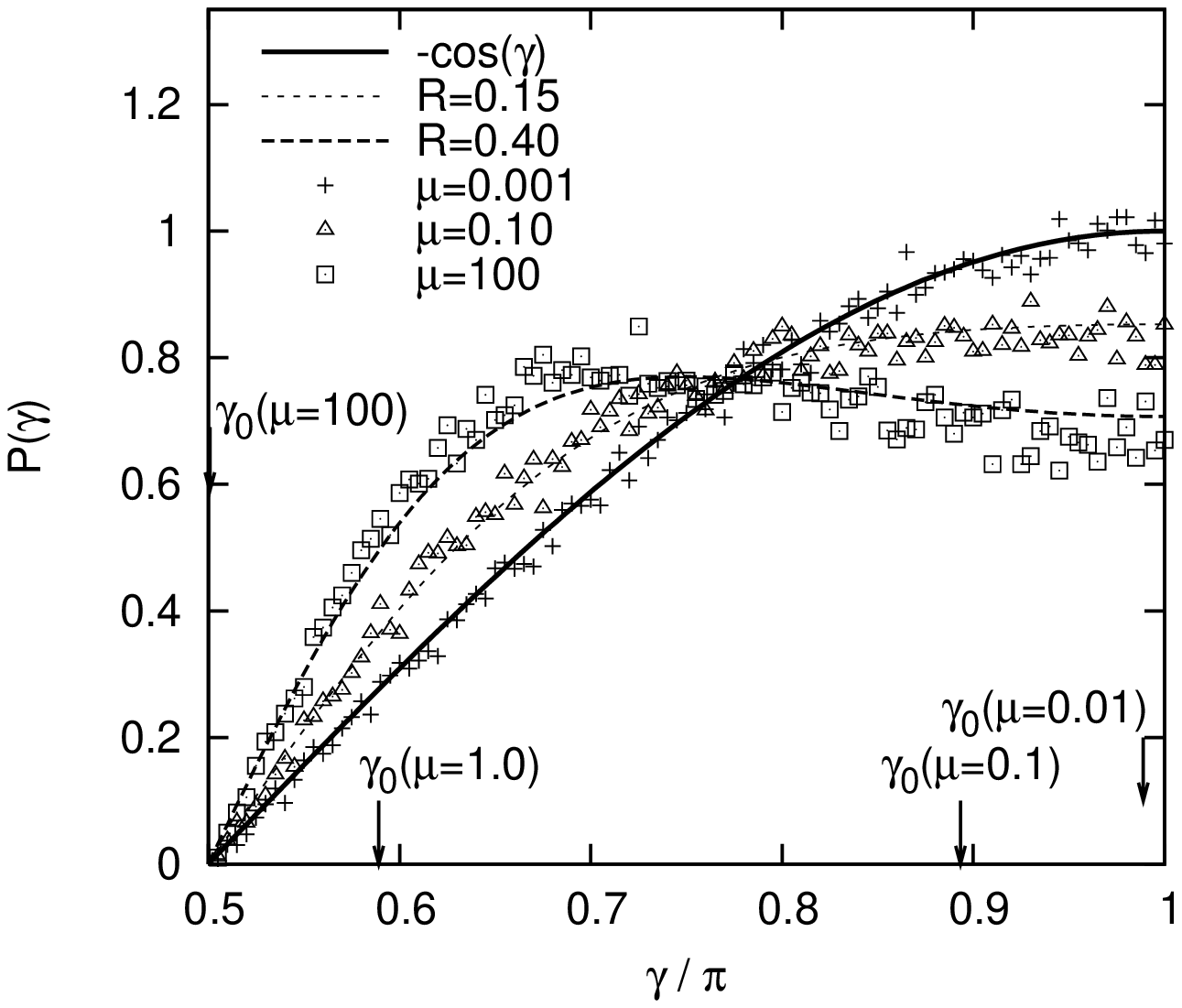,height=7.5cm,angle=-90}
\end{center}
\caption{Plots of the probability distribution of $\gamma$
  from simulations (symbols) and from Eq.\ (\ref{eq:Pgamma}) with $R$
  values from the simulations.  The arrows indicate the corresponding
  $\gamma_0$, while the parameters are (a) $r=0.95$, $\mu=0.5$ and
  variable $r_t^m$, and (b) $r=0.95$, $r_t^m=0.4$ and variable $\mu$.}
\label{fig:Pgamma}
\end{figure}

The assumption of ``molecular chaos'' implies a homogeneous
distribution of the collision parameter $b=2a \sin \gamma_{12}$ which
is simply related to the angle $\gamma_{12}$ between the relative
translational velocity $\vec{v}_{12}$ and the contact normal $\vec{n}$
according to $\cos \gamma_{12}=\vec{v}_{12} \cdot
\vec{n}/|\vec{v}_{12}|$, see Fig.\ \ref{fig:2pcoll}. Hence the
probability distribution of $\sin\gamma_{12}$ is constant,
$P'_{12}(\sin \gamma_{12})\equiv 1$. (The ``prime'' indicates
probability functions of the sine or the cosine of the angle.) A
uniform probability $P'$ implies for the distribution of the angle $
P_{12}(\gamma_{12})= - \cos \gamma_{12}$, so that grazing contacts
appear less probable than central collisions when a fixed interval
${\rm d}\gamma_{12}$ is considered. The uniform $P'_{12}(\sin
\gamma_{12})$ is in agreement with our numerical data, see Fig.\ 
\ref{fig:Pgamma}.

In general, the impact angle $\gamma$ between the relative velocity of
the contact point $\vec{g}$ and the contact normal $\vec{n}$ is {\it
  different} from the angle $\gamma_{12}$ between the relative
translational velocity $\vec{v}_{12}$ and the contact normal
$\vec{n}$, as displayed in Fig.\ \ref{fig:2pcoll}. The two angles are
identical only in the case of smooth particles or in the limit of
vanishingly small rotational velocities. In the general case we
compute $P'(\cos \gamma)$ by averaging over all binary collisions
\begin{equation}
P'(\cos \gamma)=\left \langle \delta 
                     \left (\cos \gamma-\frac{\vec{g} \cdot \vec{n}}
                            {|\vec{g}|} \right ) 
                \right \rangle_{coll} ~.
\end{equation}
This average can only be computed approximately. We assume that the
translational and rotational velocities of the colliding particles are
distributed according to Gaussians with a temperature $T_{\rm tr}$ for
the translational and a temperature $T_{\rm rot}$ for the rotational
velocities. Within this approximation the above average is given
explicitly by
\begin{equation}
P'(\cos \gamma)=\frac{J \left ( \delta \left (
                                \cos \gamma-\frac{\vec{g} \cdot \vec{n}}
                     {|\vec{g}|} \right ) \right ) }{J(1)}
\end{equation}
with the phase space integral
\begin{equation}
J(X)=\int d\Gamma_{1} d\Gamma_{2} \ (\vec{v}_{12} \cdot \vec{n}) \
\Theta(-\vec{v}_{12} \cdot \vec{n})\ \delta(|\vec{r}_{12}|-2a) \ 
X ~,\nonumber
\end{equation}
where $X=X(\Gamma_{1},\Gamma_{2})$, and the phase space element
\begin{equation}
 d\Gamma_{k}=d^2r_k d^2v_k d\omega_k e^{-m v_{k}^2/(2 T_{\rm tr})}
 e^{-I w_{k}^2/(2 T_{\rm rot})}\nonumber
\end{equation}
for $k=1, 2$.

The remaining integrals can be computed analytically, yielding the
following expression for the impact angle distribution
\begin{equation}
P(\gamma) = - \frac{\left( 1+R/q \right) \cos{\gamma}}
                 { \left( 1 + [R/q]\cos^2{\gamma} \right)^{3/2} }~.
\label{eq:Pgamma}
\end{equation}
Here we have introduced the ratio of rotational and translational
temperatures $R:=T_{\rm rot}/T_{\rm tr}$ and recall $q=I/(ma^2)$.  The
probability distribution $P(\gamma)$ is compared to the results of our
simulations in Fig.\ \ref{fig:Pgamma}; reasonably good agreement is
observed. With increasing rotational velocities, contacts with large
$g_t$ (small $\gamma$) become more and more frequent due to the
increasing rotational contribution.  On the other hand, collisions
with vanishing $g_t$ (large $\gamma$) become less probable, since the
rotational contribution leads to a net increase of $g_t$.

\section{Differential Equations in Mean Field Theory Approximations}
\label{sec:DGL.Mean.Field}

In the following we present different approximations for frictional
particles, referred to as models A-E.  Model A is the well known model
using constant coefficients of normal and tangential restitution, cf.,
e.g., \cite{jenkins85b,huthmann97}.  Model E implements Coulomb
friction as introduced by Walton \cite{walton86}.  While model A is
the mean field solution for rough particles with a constant
coefficient of tangential restitution, model E is the mean field
solution for particles with Coulomb friction.  Models B through D are
approximations to model E that may be simpler to deal with but have
significant shortcomings.

The starting point of our mean-field approach is the theory of Ref.\ 
\cite{huthmann97} for a freely cooling gas of rough particles with a
constant coefficient of tangential restitution ($r_t={\rm } const.$,
corresponding to the limit $\mu \to \infty$).  The theory is based on
a pseudo--Liouville--operator formalism and on the assumption of (i) a
homogeneous state, (ii) independent Gaussian probability distributions
of all degrees of freedom, i.e.\ all components of the translational
and the rotational velocities, and (iii) the assumption of ``molecular
chaos'', i.e.\ subsequent collisions are uncorrelated.  The agreement
with simulations is very good as long as the above assumptions are
valid \cite{luding98d}.

The main outcome of this approach is a set of coupled time evolution
equations for the translational and rotational MF temperatures $T_{\rm
  tr}$ and $T_{\rm rot}$ \cite{huthmann97} which can be extended to
also describe arbitrary energy input (driving)
\cite{cafiero00,cafiero02,luding03c}.  Given the random driving
temperature $T_{\rm dr}$ and an energy input rate $f_{\rm dr}$, as
defined above, one just has to add the positive rate of change of
translational energy $H_{\rm dr}$, see Eq.\ (\ref{eq:Hdr}), to the
system of equations \cite{cafiero00}.

\subsection{Model A: Constant tangential restitution $r_t=r_t^m$}
\label{sec:alteErgebnisse}

We recall the results of the mean field theory for the model with a
constant coefficient of tangential restitution which is obtained from
the general case in the limit $\mu \rightarrow \infty$ (see Eqs.\ (14)
in Ref.\ \cite{luding98d}). The system of coupled equations reads in
2D:
\begin{equation}\label{eq:einfacheDGLs}
\begin{array}{lll}
  \frac{d}{dt}  T_{\rm tr}(t) &= H_{\rm dr} ~ + & \,\,\, G \left[
  - A T_{\rm tr}^{3/2} + B T_{\rm tr}^{1/2} T_{\rm rot}\right] \,,\\
  \frac{d}{dt}  T_{\rm rot}(t) &= & 2 G \left[
  B' T_{\rm tr}^{3/2} - C T_ {\rm tr}^{1/2} T_{\rm rot}\right] \,.
\end{array}
\end{equation}
Note the choice of signs which lead to positive coefficients.  Based
on more physical arguments, $A$ quantifies the dissipation of
translational energy, $B$ and $B'$ correspond to the interchange of
energy between the translational and rotational degrees of freedom,
and $C$ describes the dissipation of rotational energy. The
coefficient $G$ sets the time-scale of the system, i.e.\ the collision
rate (per particle) $\tau^{-1}=(1/2) G T_{\rm tr}^{1/2}$, with
\begin{equation}\label{eq:G}
G = \frac{8}{a\sqrt{\pi m}}~\nu~g_{2a}(\nu) ~.
\end{equation}
Here $g_{2a}(\nu)$ denotes the pair correlation function at contact.
In the approximation proposed by Henderson
\cite{henderson75,henderson77,verlet82,jenkins85b,sunthar99},
$g_{2a}(\nu)=(1-7\nu/16)/(1-\nu)^2$, it depends only on the 2D volume
fraction of the granular gas $\nu=\pi a^2 N/V$. The four constants
$A$, $B$, $B'$ and $C$ read in this limit
\begin{eqnarray}
\label{eq:Ainf}
         A = A_r + A_{\eta_0}~, \quad  A_r &:=& \frac{1-r^2}{4}~,\\ 
         A_{\eta_0} &:=& \frac{\eta_0}{2}(1-\eta_0)~,\\
\label{eq:Binf}
 B' = B = B_{\eta_0} &:=& \frac{\eta_0^2}{2q}~, {\rm ~~and~~}               \\
\label{eq:Cinf}
         C = C_{\eta_0} &:=& \frac{\eta_0}{2q} 
\left (1-\frac{\eta_0}{q}\right ) ~.
\end{eqnarray}
It is useful to define a function 
\begin{equation}
\eta(r_t):=\frac{q(1+r_t)}{2(q+1)}, 
 {\rm~for~} 0 \le \eta(r_t) \le \frac{q}{q+1} < 1 ~,
\label{eq:eta}
\end{equation}
which has to be evaluated at constant tangential restitution
$r_t=r_t^m$ in the limit $\mu \to \infty$
\begin{equation}
 \eta_0:=\eta(r_t^m)=\frac{q (1+r_t^m)}{2(q+1)}~.
\label{eq:eta0}
\end{equation}

\subsection{Model B: Simplified mean tangential restitution 
$r_t=\langle r_t \rangle_{12}$}\label{sec:simple.mean}

A first step beyond the above theory with a constant
$\eta_0=\eta(r_t^m)$, is the replacement of $r_t(\gamma)$ by its
average
\begin{eqnarray}
\label{eq:average_r_t_allgemein}
\avg{r_t}_{} &= \int\limits_{\pi/2}\limits^{\pi} d \gamma \,
P(\gamma) r_t(\gamma)~. 
\end{eqnarray}
The integral over $\gamma_{}$ from $\pi/2$ to $\pi$, has to be split
into two parts, one corresponding to the range $\pi/2
<\gamma_{}<\gamma_0$ for which there is Coulomb sliding with $r_t$
given by Eq.\ (\ref{eq:betagamma.nurCoulomb}), and a second part
corresponding to the range $\gamma_0 \le \gamma \le \pi$, for which
there is sticking with constant $r_t=r_t^m$ (see Fig.\ \ref{fig:r_t}).
The critical angle $\gamma_0$ is given by
\begin{equation}\label{eq:c}
c: = -\cot \gamma_0=\frac{q(1+r_t^m)}{\mu(q+1)(1+r)} >0~.
\end{equation}

To simplify the computation, we use the approximation $P(\gamma)
\approx P_{12}(\gamma) = - \cos{(\gamma)}$, such that
\begin{eqnarray}
\label{eq:average_r_t_12}
\avg{r_t}_{12} &=
-1 + \frac{q+1}{q} (1+r) \mu~ \ln \left( c+f \right )~.
\end{eqnarray}
with the abbreviation
\begin{equation}
 f:=\sqrt{1+c^2}\,.
\label{eq:f}
\end{equation}
The averaged coefficient of tangential restitution $\avg{r_t}_{12}$
must be inserted into $\eta$ in Eq.\ (\ref{eq:eta}). Thus we obtain
the same set of coefficients as in Eqs.\ 
(\ref{eq:Ainf})-(\ref{eq:Cinf}) with $\eta_0$ replaced by
\begin{equation}
\eta_1:= \eta(\avg{r_t}_{12})=\frac{\eta_0}{c} \ln
\left( c+f \right ) ~.
\label{eq:eta1}
\end{equation}

In this approach, only the average value of $r_t$ is considered and
fluctuations of $r_t$ with $\gamma$ are neglected. Furthermore the
difference between $\gamma$ and $\gamma_{12}$ has been ignored in the
averaging procedure. In contrast to model A this is the simplest model
to incorporate the coefficient of Coulomb friction $\mu$,
$\avg{r_t}_{12}=\avg{r_t}_{12}(\mu)$.

\subsection{Model C: Mean tangential restitution 
  $r_t=\langle r_t \rangle_{}(R)$}
\label{sec:full.mean}

In model C we again replace $r_t(\gamma)$ by its average but use the
correct impact angle probability distribution function $P(\gamma)$ from
Eq.\ (\ref{eq:Pgamma}) in the averaging procedure.  The result is an
$R$-dependent averaged coefficient of tangential restitution

\begin{eqnarray}
\label{eq:average_r_t}
& &\avg{r_t}_{}(R)  = -1+\frac{q+1}{q} \, 
                         \frac{1+r}{4} \frac{\mu}{x} \times \\
& & \ln{\left\{ \frac{ \frac{R}{q}
 (f-c)^2(x\tilde{f}-f+c\frac{R}{q})}
    {(x\tilde{f}-f-c\frac{R}{q})^2(x\tilde{f}+f-c\frac{R}{q})}\right\}} \nonumber      
\end{eqnarray}
with
\begin{equation}
x^2 \equiv x^2(R) := 1+R/q\,,
\lab{eq:x}
\end{equation}
\begin{equation}
\label{eq:ff} 
{\tilde f} \equiv {\tilde f}(R) := \sqrt{1+x^2c^2}\,,
\end{equation}
and $f$ defined in Eq.\ (\ref{eq:f}).  Note that $x$ is an implicit
function of time through $R$.  For $R\to 0$ ($x\to 1$) Eq.\ 
(\ref{eq:average_r_t}) reduces to Eq.\ (\ref{eq:average_r_t_12}) -- as
expected.  For $R\to\infty$ ($x\to\infty$) there is no friction and
$\avg{r_t}(R)\to-1$.

We formally get the same differential equations
(\ref{eq:einfacheDGLs}) but with non-constant coefficients $A=A(R)$,
$B'=B=B(R)$, and $C=C(R)$ which are obtained by replacing $\eta_0$ by
$\eta(\avg{r_t}_{}(R))$ in Eqs.\ (\ref{eq:Ainf})-(\ref{eq:Cinf}).
These coefficients are implicitly time dependent via $R$.

\subsubsection{Constant tangential restitution limit}
\label{sec:full.mean.const.rest}

In the limit $\mu \rightarrow \infty$, $c\to 0$.  In that case model C
reduces to model A.

\subsubsection{Weak friction limit}
\label{sec:full.mean.weak.friction}

For $\mu\to 0$, $c\to \infty$ we recover smooth spheres with
$\avg{r_t}_{}\to -1$.  A series expansion to lowest order in $\mu$
(equivalent to lowest order in $c^{-1}$) of Eq.\ 
(\ref{eq:average_r_t}) reads
\begin{align}\nonumber
  \avg{r_t}_{}(R) &= -1+\frac{q+1}{q}(1+r)\frac{\mu}{x}
  {\Big{\{}}|\ln{\left(\mu\right)}| + \ln{(x)}
  +\\\label{eq:average_r_t.small.mu} & \qquad\qquad
  \ln{\left(\frac{2\eta_0}{1+r}\right)}{\Big{\}}}+\mathcal{O}{(\mu^3)}~,
\end{align}
expressed in terms of $x$ and $\mu$.

As long as $x$ stays finite (which is the case for a driven system)
the leading order is thus $\mu |\ln{(\mu)}|$ for small $\mu$.  For $x
\to 1$, Eq.\ (\ref{eq:average_r_t.small.mu}) yields the same result as
Eq.\ (\ref{eq:average_r_t_12}) in leading order in $c^{-1}$.

\subsubsection{Comparison of model B and model C}

\begin{figure}[htb]
\begin{center}
\epsfig{file=\picdirectory/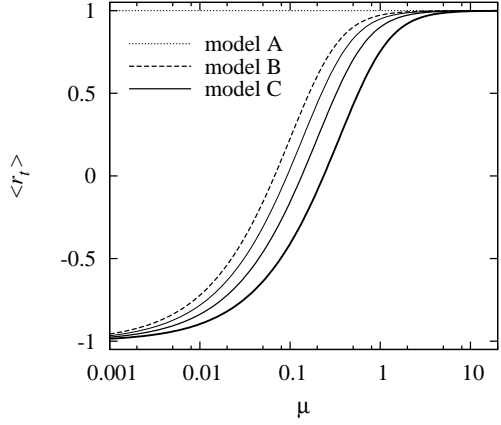,height=7.2cm,angle=-90}
\end{center}
\caption{Expected mean tangential restitution, $\langle r_t \rangle$, 
  as function of the friction coefficient $\mu$ for models A, B, and
  C.  The parameters used are $r=0.95$, $r_t^m=1.0$ (for A, B) and
  different $R=1.0$, $0.40$, and $0.15$ (model C: solid lines from
  right to left).  }
\label{fig:rt_av}
\end{figure}

Due to the implicit nature of model C it is rather difficult to work
out its predictions, e.g., for the ratio of temperatures.  Therefore,
we present here the mean tangential restitution from models A, B, and
C in Fig.\ \ref{fig:rt_av}. Note that $\langle r_t \rangle$ for model
C depends not only explicitly on $\mu$ but also implicitly through
$R$. To keep the discussion simple, we present results only for some
constant, representative values of $R$.  The mean restitution for
large $R$ is smaller (or equivalently, the corresponding $\mu$ is
larger) than for small $R$.  Models B and C become indistinguishable
in the limit $R \to 0$, as expected.

\subsection{Model D: Variable (simplified) tangential restitution
  $r_t(\gamma_{12})$}
\label{sec:simplified.integrals}

In this section and in the following one, we discuss a coefficient of
tangential restitution which depends on $\gamma$. Model D is defined
by approximating $\gamma \approx \gamma_{12}$, which is strictly true
only for $R \to 0$ or $\mu \to 0$ (or equivalently $r_t^m\to-1$).  We
again obtain the same differential equations (\ref{eq:einfacheDGLs})
for $T_{\rm rot}$ and $T_{\rm tr}$ with the coefficients
 \begin{eqnarray}\label{eq:const2dnew} 
 A~ = A _{\mu} & = & A_r+\left [A_{\eta_0} + A^*\right ]/f^3\,, \\ \nonumber
 B~ = B _{\mu} & = & \left [ B_{\eta_0} + B^* \right ] / f^{}\,,    \\  \nonumber
 B' = B'_{\mu} & = & \left [ B_{\eta_0} + B'^* \right ] / f^3\,,  \\ \nonumber
 C~ = C_{\mu} & = & \left [ C_{\eta_0} + C^* \right ] / f^{} \,,     \nonumber
 \end{eqnarray}
 and $f$ defined in Eq.\ (\ref{eq:f}). The terms that originate from
 Coulomb sliding are denoted by an asterisk and are given explicitly
 by
\begin{eqnarray}
\label{eq:const2dnew2} 
 B^*  
& = &\frac{\eta_0^2 c^2 }{2q(f+1)^2} ~,                  \\
B'^* 
& = & (2f+1) B^* ~,
                                                              \nonumber \\
A^* 
& = &      \eta_0 c^2/2-q B^{*}
                                               ~, {\rm~and~}  \nonumber \\
C^* 
& = &       \left ( \eta_1 f - \eta_0 - 2 B^*
      \right )/(2q) ~,\nonumber
\end{eqnarray}
expressed in terms of $f$ [cf. Eq.\ \req{eq:f}], $\eta_0$ [cf. Eq.\ 
\req{eq:eta0}], $\eta_1$ [cf.  Eq.\ \req{eq:eta1}], and $q$ [cf. Eq.\ 
\req{eq:q}].  The terms $B^*$ and $B'^*$ are strictly positive, while
the dissipation correction terms $A^*$ and $C^*$, in principle, can
change sign.  Note also that $B^*$ and $B'^*$ are not identical here.
All coefficients depend on the system parameters only.  They are
constants in time -- in contrast to model C (and E as will be shown
later).

 \subsubsection{Constant tangential restitution limit}
\label{sec:simplified.integrals.const.rest}
 
In the limit $\mu \rightarrow \infty$, one has $c \rightarrow 0$,
i.e.\ $f \rightarrow 1$, and all correction terms
$\{A^*,B^*,B'^*,C^*\} \rightarrow 0$ so that one obtains Eqs.\ 
(\ref{eq:einfacheDGLs})-(\ref{eq:Cinf}).  Note in particular that the
coefficients $B_\mu$ and $B'_\mu$ are equal only in the limit $\mu \to
\infty$ \cite{huthmann97}.

 \subsubsection{Weak friction limit}
\label{sec:simplified.integrals.weak.friction}
 
In the limit $\mu \rightarrow 0$ ($c\to\infty$, $f\to c$), the lowest
order expansion in $c^{-1}$ leads to an approximation of the
coefficients in Eqs.\ (\ref{eq:const2dnew}), where we have used
$\eta_0/c = \mu (1+r)/2$:
 \begin{eqnarray}
 \label{eq:smallmu}
 B_\mu & = & 
 \frac{\eta_0}{q}\ \frac{1+r}{2}\mu + \mathcal{O}{\left(\mu^{2}\right)}\,,\\
 \nonumber 
 B'_\mu & = & \frac{1}{q} \left(\frac{1+r}{2}\right)^2\mu^2 + \mathcal{O}\left(\mu^{3}\right)\,,\\
 \nonumber 
 A_\mu & = & A_r +
\frac{1+r}{4}\mu +
\mathcal{O}\left({\mu^{2}}\right)\,,\\
 \nonumber 
 C_\mu
  & = &
  \frac{1}{2q}\frac{1+r}{2}\mu\left(|\ln{(\mu)}|+\ln{\left(\frac{4\eta_0}{1+r}\right)} -2\frac{\eta_0}{q}\right) \\\nonumber &~& \vspace*{0.1cm} ~
+ \mathcal{O}\left(\mu^{2}\right)\ .
 \end{eqnarray}
 
 From Eqs.\ (\ref{eq:smallmu}), we learn that $B'_\mu$ is second order
 in $\mu$, whereas $B_\mu$ is first order in $\mu$, reflecting an {\em
   asymmetry} in the energy transfer rates.  On the other hand, $A_\mu
 \approx A_r$ is almost constant, whereas $C_\mu$ depends on $\mu$
 logarithmically which is an artifact of our approximation
 $\gamma_{12} \sim \gamma$, see Eq.\ (\ref{eq:smallmu2}) below.

\subsection{Model E: Variable (exact) tangential restitution $r_t(\gamma)$}
\label{sec:full.version}

The final step of refinement of the MF theory is to use $r_t(\gamma)$,
instead of $r_t(\gamma_{12})$, to compute the coefficients.  This is
the full mean field theory.  The calculation is similar to the one for
3D in \cite{herbst00} and is presented in appendix
\ref{appendix:modelE}.  We obtain the following coefficients, to be
inserted into Eqs.\ (\ref{eq:einfacheDGLs}),
 \begin{eqnarray}\label{eq:const3dnew} 
 A~ = {\widetilde A}_{\mu}(R) & = &
 {\color{green}A_r}+\left[{\color{red}A_{\eta_0}} + {\widetilde
     A}^*\right] {\Big / } {\tilde f}^3 \\ \nonumber
 B~ = {\widetilde B}_{\mu}(R) & = & \left [ {\color{red}B_{\eta_0}} +
   {\widetilde B}^* \right] {\Big / } {\tilde f}^3
 \\  \nonumber
 B' = {\widetilde B}'_{\mu}(R) & = & \left[ {\color{red}B_{\eta_0}} +
   {\widetilde  B}'^* \right] {\Big / } {\tilde f}^3  
\\ \nonumber
 C~ = {\widetilde C}_{\mu}(R) & = & \left [ {\color{red}C_{\eta_0}} +
   {\widetilde C}^* \right ] {\Big / } {\tilde f}^3 ~,     \nonumber
 \end{eqnarray}
 with $\tilde f$, $x$ and $c$ defined in Eqs.\ (\ref{eq:ff}),\req{eq:x}
 and \req{eq:c}, respectively.  The new correction terms are in
 detail:
\begin{eqnarray}
\label{eq:const3dnew2} 
 {\widetilde B}^* &=& -\eta_0 c^2/(2q) ~,\\ \nonumber
 {\widetilde B}'^*  &=& \ \frac{(2{\tilde f}+1)(\eta_0 c x^2)^2}
{2q({\tilde f}+1)^2} \\ \nonumber
 {\widetilde A}^* 
 &=& - q \left ({\widetilde B}^{*} 
+ {\widetilde B}'^* \right ) ~,
{\rm ~ and ~} \\ \nonumber
 {\widetilde C}^* 
  &=& - x^2 {\widetilde B}^{*}   ~,
\end{eqnarray}
with $q$ and $x$ as introduced in Eqs.\ \req{eq:q} and \req{eq:x}.
Interestingly, we find now a negative ${\widetilde B}^*$ together with
positive coefficients ${\widetilde B}'^*$ and ${\widetilde C}^*$; only
${\widetilde A}^*$ can be both positive and negative.  Like in model C
but in contrast to models A, B, and D, here the coefficients are
implicit functions of time, again.

In conclusion, models D and E appear similar in shape but there are
several striking differences: (i) The division by $f$ and $f^3$ in
model D is in contrast with the division by $\tilde{f}^3$ in model E,
(ii) the term $B^*$ in model D is always positive, while $\tilde{B}^*$
in model E is always negative, (iii)the sign of $C^*$ in model D is
not determined a-priori, while the term $\tilde{C}^*$ is always
positive, (iv) among the correction terms of model E, only
$\tilde{B}^*$ is independent of $R$, and (v) the more refined theory
appears in a simpler form, especially the term $\tilde{C}^*$.

\subsubsection{Constant tangential restitution limit}
\label{sec:full.version.const.rest}

The limit of constant tangential restitution can be reached by taking
the limit $\mu \to \infty$.  In this case $c \to 0$, ${\tilde f} \to
1$ and thus all additional coefficients ${\widetilde A}^*$,
${\widetilde B}^*$, ${\widetilde B}'^*$, and ${\widetilde C}^*$ vanish
such that Eqs.\ (\ref{eq:einfacheDGLs})-(\ref{eq:Cinf}) are recovered.

\subsubsection{Weak friction limit}
\label{sec:full.version.weak.friction}

In the limit $\mu \rightarrow 0$ ($c\to\infty$, ${\tilde f} \to xc$)
an expansion to the lowest order in $\mu$ leads to an approximation of
the coefficients in Eqs.\ (\ref{eq:const3dnew}) when we remember that
$\eta_0/c=(1+r)\mu/2$:
\begin{eqnarray}
\label{eq:smallmu2}
{\widetilde B}_{\mu}(R) & = & 
{\color{dark-green} - \frac{1}{{\color{black}2q}x^3} \frac{1+r}{2} \mu } +
 \mathcal{O}\left(\mu^{3}\right)\,,\\ 
\nonumber
 {\widetilde B}'_{\mu}(R) & = & 
{\color{black}\frac{1}{q}} {\color{red-green}\left(\frac{1+r}{2}\right)^2\mu^2} + \mathcal{O}\left(\mu^{3}\right)\,,
\\ \nonumber
 {\widetilde A}_{\mu}(R) 
& = & {\color{green}A_r} - q \left({\color{dark-green}{\widetilde B}_{\mu}(R)}
 + {\color{red-green}{\widetilde B}'_{\mu}(R)} \right) 
 + \mathcal{O}\left(\mu^{3}\right)\,,
 \\ \nonumber
 {\widetilde C}_{\mu}(R) & = & 
{\color{blue} -x^2 {\widetilde B}_{\mu}(R)} 
+ \mathcal{O}\left(\mu^{3}\right)
~.
\end{eqnarray}
Since $x=x(R)$ approaches one in the weak friction limit, both
${\widetilde B}_{\mu}(R)$ and ${\widetilde C}_{\mu}(R)$ are
proportional to $\mu$ in leading order. To lowest order in $\mu$, Eq.\ 
(\ref{eq:smallmu2}) predicts ${\widetilde
  A}_{\mu}(R)=A_r+\mathcal{O}(\mu)$, i.e.\ proportional to $\mu^0$,
while ${\widetilde B}'_{\mu}(R)$ is proportional to $\mu^2$.

For $\mu \ll 1$, Eqs.\ (\ref{eq:einfacheDGLs}) with
(\ref{eq:smallmu2}) simplify to
\begin{eqnarray}\label{eq:tr.mu.small}
  \frac{d}{dt}T_{ {\rm tr}}(t) =& H_{\rm dr}-G
  T_{\rm tr}^{3/2} \left({\color{blue} \frac{1-r^2}{4}} 
+\mathcal{O}{\left(\mu\right)}\right)\,,
\end{eqnarray}
which means that in the limit of low friction the differential
equations for $T_{\rm tr}$ and $T_{\rm rot}$ decouple.  In the
non-driven case this leads to surviving rotational energy (not show),
similar to Refs.\ \cite{herbst00,goldhirsch04}.

\section{Steady State}
\label{sec:steadystate}

Before discussing the approach to the stationary state in the next
chapter, we first elucidate the stationary state and compare results
of our simulations to various levels of refinement of the mean field
theory.
 
\subsection{Analytical results}
\label{sec:steadystate_analytical}

By imposing $\frac{d}{dt} T^{\rm stat}_{\rm tr}\!=\!0$ and
$\frac{d}{dt} T^{\rm stat}_{\rm rot}\!=\!0$ one gets the steady state
values of the rotational and the translational temperatures.  For
models A, B and D, the coefficients in the differential equation do
not depend on $R$ (or $x$).  Therefore the solution is simply
\begin{equation}\label{eq:mfeq}
T^{\rm stat}_{\rm rot} = R^{\rm stat} T^{\rm stat}_{\rm tr}\,,{\rm ~~and~~} \,
T^{\rm stat}_{\rm tr} = \left( \frac{H_{\rm dr}}{G \mathcal{I}} \right )^{2/3} ,
\end{equation}
with 
\begin{equation}\label{eq:R_I}
R^{\rm stat}={B'}/{C} \,,{\rm ~~and~~} \mathcal{I} = A-BR^{\rm stat} ~,  
\end{equation}
as discussed in more detail for all models in the following.

\subsubsection{Model A}

For model A, the steady state ratio of rotational to translational
energies is
\begin{equation}
R^{\rm stat} = \frac{q\eta_0}{q-\eta_0}   
\label{eq:R_A}
\end{equation}
and  the energy dissipation factor is
\begin{equation}
\mathcal{I} =\frac{1-r^2}{4} - \frac{\eta_0}{2}(1-\eta_0) -
\frac{\eta_0^3}{2(q-\eta_0)} ~.
\label{eq:I_A}
\end{equation}
Note here again that model A does not contain any dependence on the
coefficient of friction $\mu$.

\subsubsection{Model B}

Model B evolves from model A, by just replacing $\eta_0$ by
$\eta_1(\mu)=(\eta_0/c) \ln{(c+f)}$ from Eq.\ (\ref{eq:eta1}) in the
above two Eqs.\ (\ref{eq:R_A}) and (\ref{eq:I_A}), so that, e.g.,
\begin{equation}
R^{\rm stat} = \frac{q \eta_1}{q-\eta_1}
           = \frac{q(\eta_0/c) \ln{(c+f)}}
                  {q-(\eta_0/c)\ln{(c+f)}} ~.      \nonumber\\
\end{equation}

In the limit of small $\mu \ll 1$, the leading order terms are $R^{\rm
  stat} \approx (1+r)\mu|\ln{\mu}|/2$ and $\mathcal{I} \approx
\frac{1-r^2}{4} + \mathcal{O}{(\mu|\ln{\mu}|)}$.

\subsubsection{Model D}

From model D, the following, more complex terms are obtained:
\begin{equation}
R^{\rm stat}_\mu=\frac{B'_\mu}{C_\mu}
              =\frac{\left[B_{\eta_0} + B'^*\right]}
                    {\left[C_{\eta_0} + C^* \right]}
               \frac{1}{f^2}
\underset{\mu \ll 1}{\approx} (1+r) \mu |\ln{\mu}|
 \nonumber
\end{equation}
and
\begin{equation}
\mathcal{I}_\mu = A_\mu - B_\mu R^{\rm stat}_\mu 
  \underset{\mu \ll 1}{\approx} \frac{1-r^2}{4} + \mathcal{O}{(\mu)} ~,
\nonumber
\end{equation}
so that, asymptotically for $\mu \ll 1$, model D leads to 
behavior similar to that of model B.

\subsubsection{Model E}

Formally, we can write down Eqs.\ \req{eq:mfeq} for model E, too.
Instead of using Eqs.\ \req{eq:R_I}, $R^{\rm stat}$ must be extracted
(numerically) from Eq.\ \req{eq:fullsolution.rot} where the left hand
side vanishes in the stationary case.  It can be show analytically
that there is always a unique solution -- in contrast to the freely
cooling case \cite{herbst00}.  With the solution for $R^{\rm stat}$ at
hand, Eq.\ \req{eq:fullsolution.tr} (with a vanishing left hand side)
can be written in the form $T^{\rm stat}_{\rm tr} = \left(
  \frac{H_{\rm dr}}{G \mathcal{I}} \right )^{2/3}$ again where
$\mathcal{I}$ is a nonlinear function of $R^{\rm stat}$ whose
particular form can be easily seen from Eq.\ \req{eq:fullsolution.tr}.

\subsubsection{Models C and E for small $\mu$}

For models C and E the coefficients in the differential equations {\it
  do} depend on $R$, so that the steady state values have to be
computed numerically for a general choice of parameters.  Analytical
results can only be achieved in the limit $\mu \ll 1$, where we can
use the expansions of the coefficients introduced in sections\ 
\ref{sec:full.mean} and \ref{sec:full.version}.

For model C we obtain to lowest order in $\mu$, the dissipation
factor $\mathcal{I} \approx A_{r}$ and, using $\eta_2=(\eta_0/2) \ln{c}$,
\begin{equation}
\begin{split}
  R^{\rm stat} &\underset{\mu \ll 1}{\approx} \frac{2\eta_2}{c}
  \sqrt{1+\left ( \frac{\eta_2}{c q} \right )^2 } +
  \frac{2}{q}\left(\frac{\eta_2}{c}\right)^2 \underset{\mu \ll
    1}{\approx} \frac{1+r}{2}\mu|\ln{\left(\mu\right)}|
  ~.\label{eq:mfeq2}
\end{split}
\end{equation}

For model E we find again $\mathcal{I} \approx A_{r}$ and
\begin{equation}\label{eq:mfeq3}
R^{\rm stat} \underset{\mu \ll 1}{\approx} 
          \frac{2 \eta_0}{c} \sqrt{1+\left ( \frac{\eta_0}{c q} \right )^2}
                    + \frac{2}{q} \left(\frac{\eta_0}{c}\right)^2 
           \underset{\mu \ll 1}{\approx} (1+r)\mu~,
\end{equation}
very similar in shape to the result from model C, besides the
logarithm $\ln c$ that is hidden in the definition of $\eta_2$.  This
leads to the qualitative difference in asymptotic behavior between
models C and E: The correct asymptotic behavior for small $\mu$ is
$R^{\rm stat} \propto \mu$.  Note again that the more refined model E
leads to a simpler analytical result than the approximated model C.

\subsubsection{Discussion}

The expansions for small $\mu \ll 1$ show that the result for $R^{\rm
  stat}$ based on model E, see Eq.\ (\ref{eq:mfeq3}), disagrees with
all other models. In model E we find that $R^{\rm stat}$ vanishes
linearly as $\mu \to 0$, whereas models A-D predict a slower decrease,
encoded in the $\mu |\ln{\mu}|$ dependence. Models A and B have the
same analytical form for $R^{\rm stat}$ if expressed in terms of
$\eta_0$ for model A and in terms of $\eta_1$ for model B. Similarly,
models C and E have the same functional dependence on $\eta$, if
$\eta_2$ is used for model C and $\eta_0$ for model E.  The comparison
of the models for arbitrary values of $\mu$ will be given in the next
subsection, where we also present the results of our simulations and
compare them to the predictions of the various mean field models.

\subsection{Comparison with simulations}

In this subsection, the steady state predictions from our models are
confronted with the numerical simulation results. Note that we present
results for rather high densities and dissipation, where our
assumptions about homogeneity of the system and the Gaussian shape of
the velocity distributions is not strictly true anymore.  However, we
want to stress the point that the present theory is astonishingly
close to the numerical simulation with experimentally relevant
parameters even when the most basic assumptions are somewhat
questionable.

\subsubsection{Variation of $r_t^m$}

\begin{figure}[htb]
\begin{center}
  \epsfig{file=\picdirectory/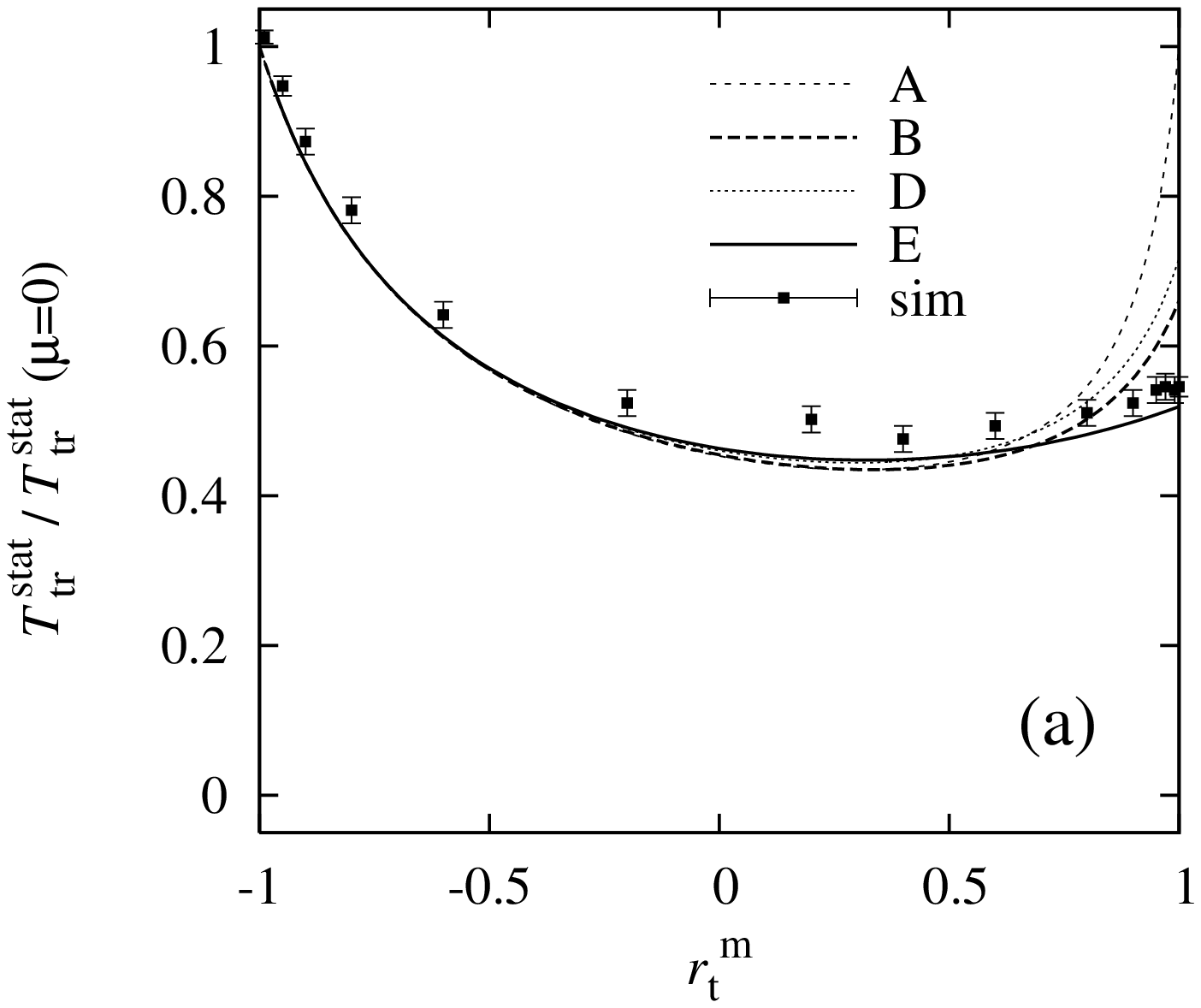,height=6.9cm,angle=-90} \\
  \epsfig{file=\picdirectory/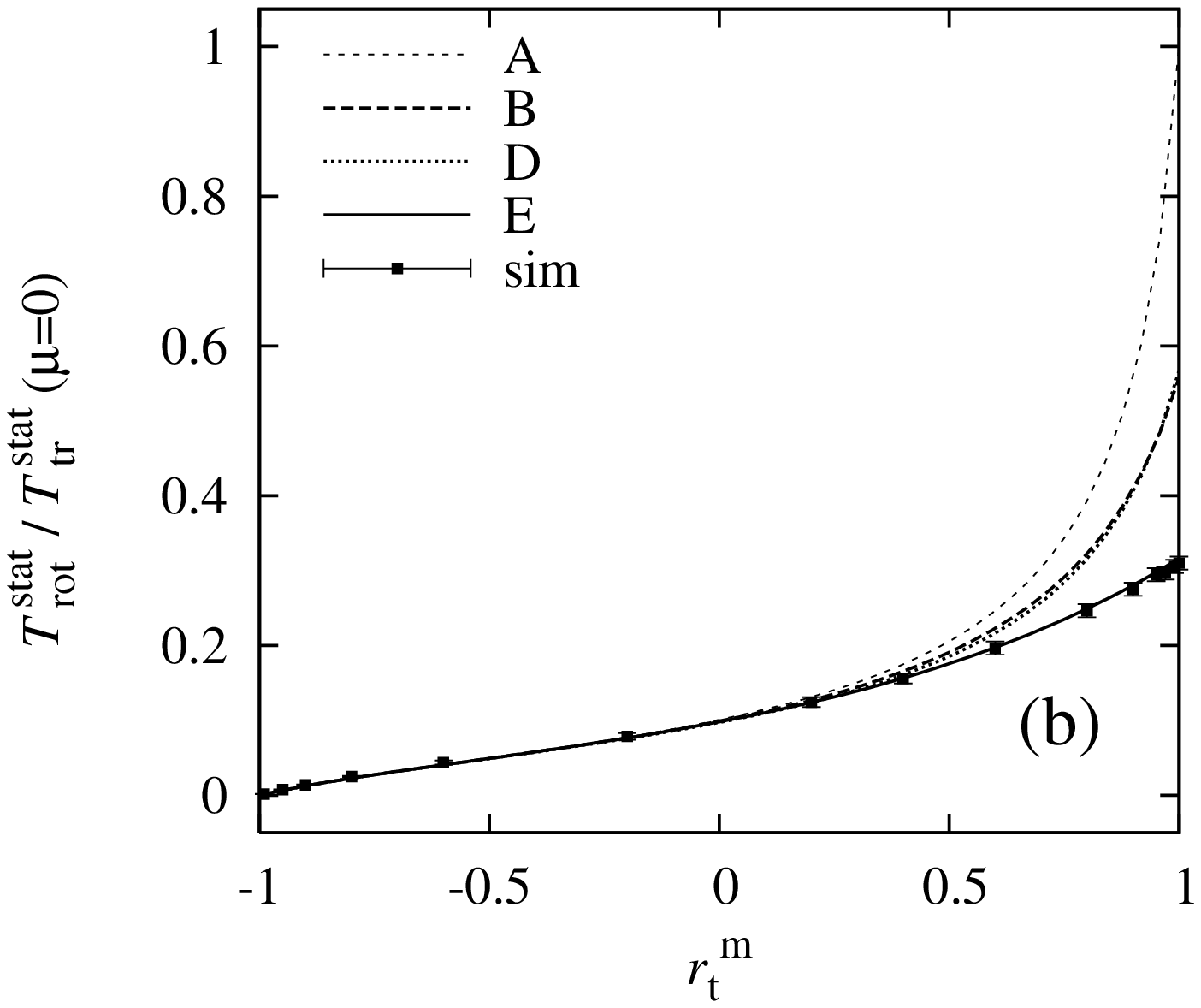,height=6.9cm,angle=-90} \\
  \epsfig{file=\picdirectory/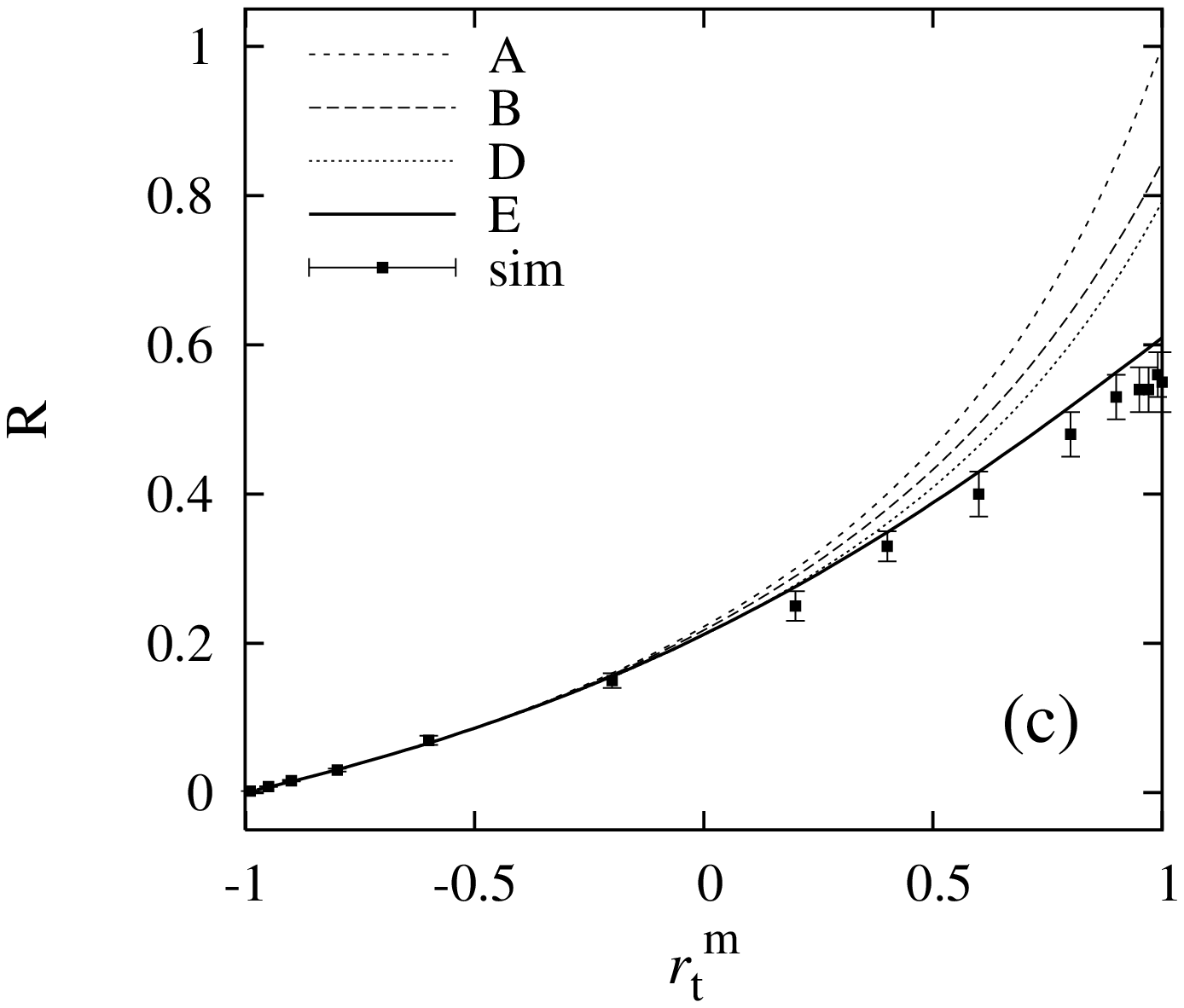,height=6.9cm,angle=-90} \\
\end{center}
\caption{Simulation results (symbols) and theory (lines) for
  the parameters $\nu=0.34$, $N=11025$, $r=0.95$, and $\mu=0.5$,
  plotted against the maximum tangential restitution $r_t^m$.  (a)
  Translational temperature $T^{\rm stat}_{\rm tr}$, and (b)
  rotational temperature $T^{\rm stat}_{\rm rot}$, plotted against
  $r_t^m$, and scaled by $T^{\rm stat}_{\rm tr}(\mu=0)$, the mean
  field value for smooth particles.  (c) Ratio of rotational and
  translational temperature $R$, plotted against $r_t^m$.  }
\label{fig:varyrtm}
\end{figure}

In Figs.\ \ref{fig:varyrtm}\ (a-c), the stationary rotational and
translational temperatures and their ratio $R$ are compared for
$r=0.95$, $\mu=0.5$ and different values of $r_t^m$; note that the
data in (a) and (b) are scaled with the expression for $\mu=0$.  The
symbols correspond to simulation data, with the error bars showing the
standard deviation from the mean values.  The lines correspond to
different refinements of the theoretical approaches, i.e.\ models A,
B, D, and E.

For $r_t^m \approx -1$, the simulations agree with all theoretical
predictions; for $r_t^m \approx 1$, large discrepancies are evident.
The more refined a model used, the better the quality of agreement.
The qualitative behavior of the data is best captured by model E, and
we relate the remaining quantitative deviations to the fact that the
simulations involve rather high density $\nu$ and comparatively strong
dissipation $r$.

\subsubsection{Variation of $\mu$ -- translational temperature}

\begin{figure}[hbt]
\begin{center}
\epsfig{file=\picdirectory/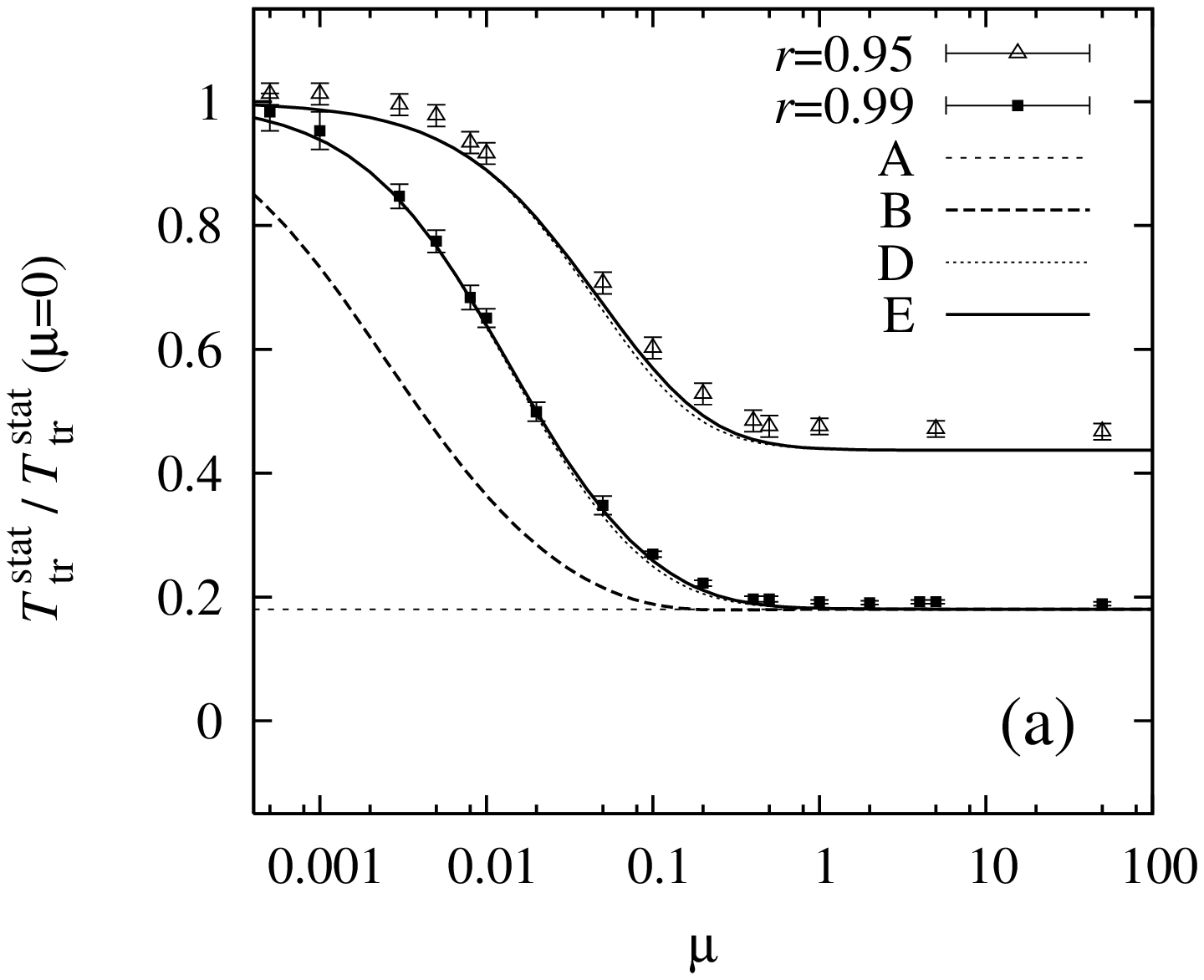,height=7.5cm,angle=-90}\\
\epsfig{file=\picdirectory/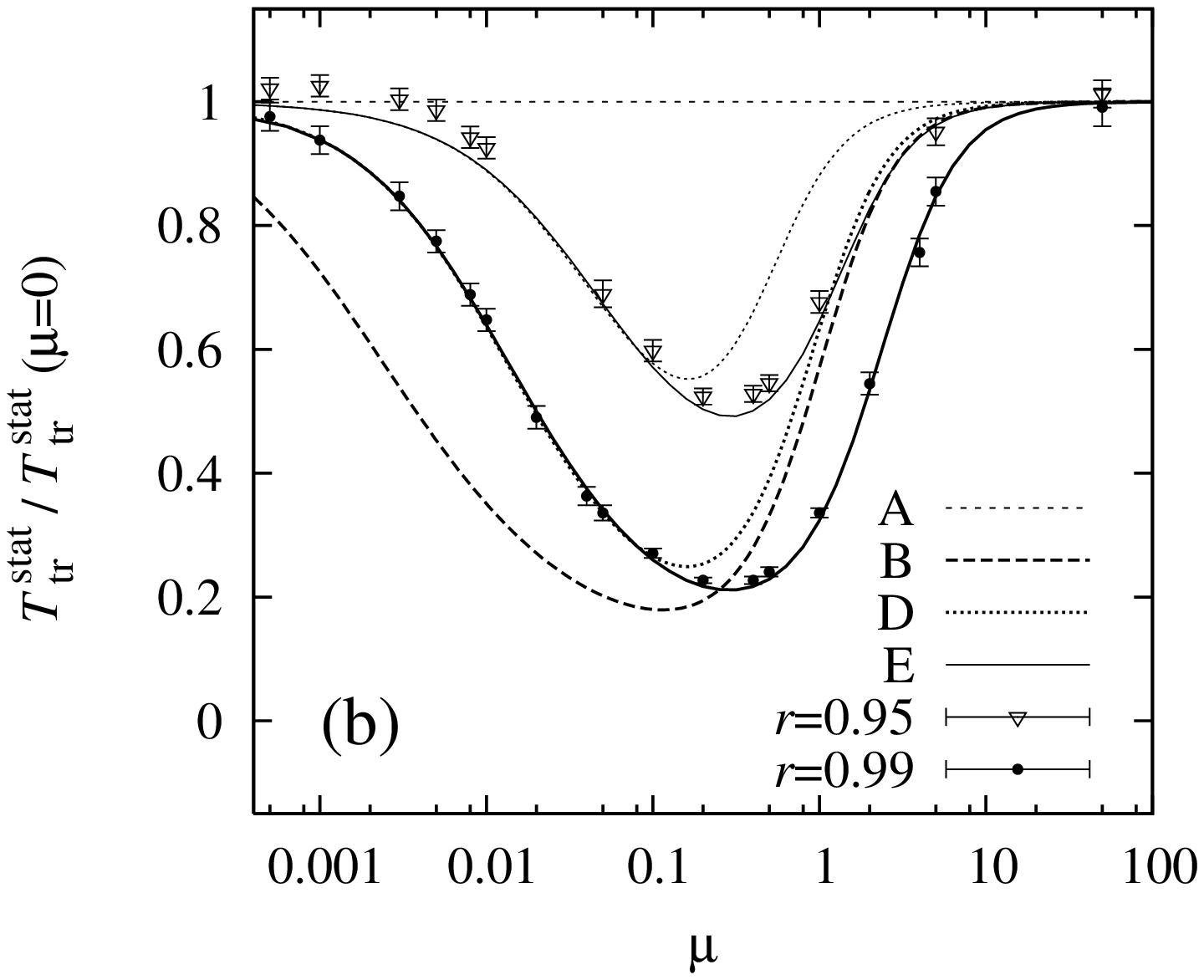,height=7.5cm,angle=-90}\\
\end{center}
\caption{Translational temper\-ature $T^{\rm stat}_{\rm tr}$ scaled by
  the mean field value for smooth particles $T^{\rm stat}_{\rm
    tr}(\mu=0)$, plotted against $\mu$, for the parameters as in Fig.\ 
  \protect{\ref{fig:varyrtm}}.  The tangential restitution
  coefficients are fixed to (a) $r_t^m=0.4$, and (b) $r_t^m=1.0$.
  Data with normal restitution $r=0.99$ (solid symbols and thick
  lines) and $r=0.95$ (open symbols and thin lines) are compared.  }
\label{fig:Tmu}
\end{figure}

In Fig.\ \ref{fig:Tmu} we plot the translational temperature in the
same way as in Fig.\ \ref{fig:varyrtm}(a), but now, we keep the values
$r_t^m=0.4$ (a) and $r_t^m=1.0$ (b) fixed and vary $\mu$.
Furthermore, we compare data for $r=0.99$ and $r=0.95$ in one plot and
observe satisfactory agreement between simulation results and the full
mean field theory, model E.  (The predictions from models A and B are
only shown for $r=0.99$.)

For (realistic) values of $r_t^m=0.4$, see Fig.\ \ref{fig:Tmu}(a), one
obtains a transition from the $\mu=0$ limit to the $\mu \to \infty$
value of the kinetic energy, over three orders of magnitude in $\mu$,
whereas for $r_t^m=1.0$, see Fig.\ \ref{fig:Tmu}(b), the kinetic
energy first decays with $\mu$ but then increases again to the
stationary state temperature of smooth particles, since no energy is
dissipated due to tangential friction for $\mu\to\infty$ and
$r_t^m=1.0$.

Here, we remark that model A, with $r_t=r_t^m$ and the limit
$\mu\to\infty$ is inadequate to model the $\mu$-dependency of the
data, it only gives the $\mu\to\infty$ limit, as expected.  Approach B
only shows qualitative agreement with our simulation data, whereas
theory D shows good quantitative agreement for small $\mu$.  The
agreement seems better for weak normal dissipation $r=0.99$, as
compared to the cases with $r=0.95$.  The deviations between
simulations and model D in the intermediate range of $\mu$ are due to
values of $R$ of the order of unity, for which the assumption
$\gamma_{12} \approx \gamma$ is {\em not} true, as pointed out above.

For weaker normal dissipation $r$, one obtains a stronger reduction of
the translational temperature in the range of strongest total
dissipation (around $\mu \approx 0.4$).  This is due to the
comparatively stronger contribution of tangential dissipation.
However, as in the previous subsection, the agreement between
simulations and model E is satisfactory, especially for $r \to 1$.

\subsubsection{Variation of $\mu$ -- rotational temperature}

\begin{figure}[htb]
\begin{center}
  \epsfig{file=\picdirectory/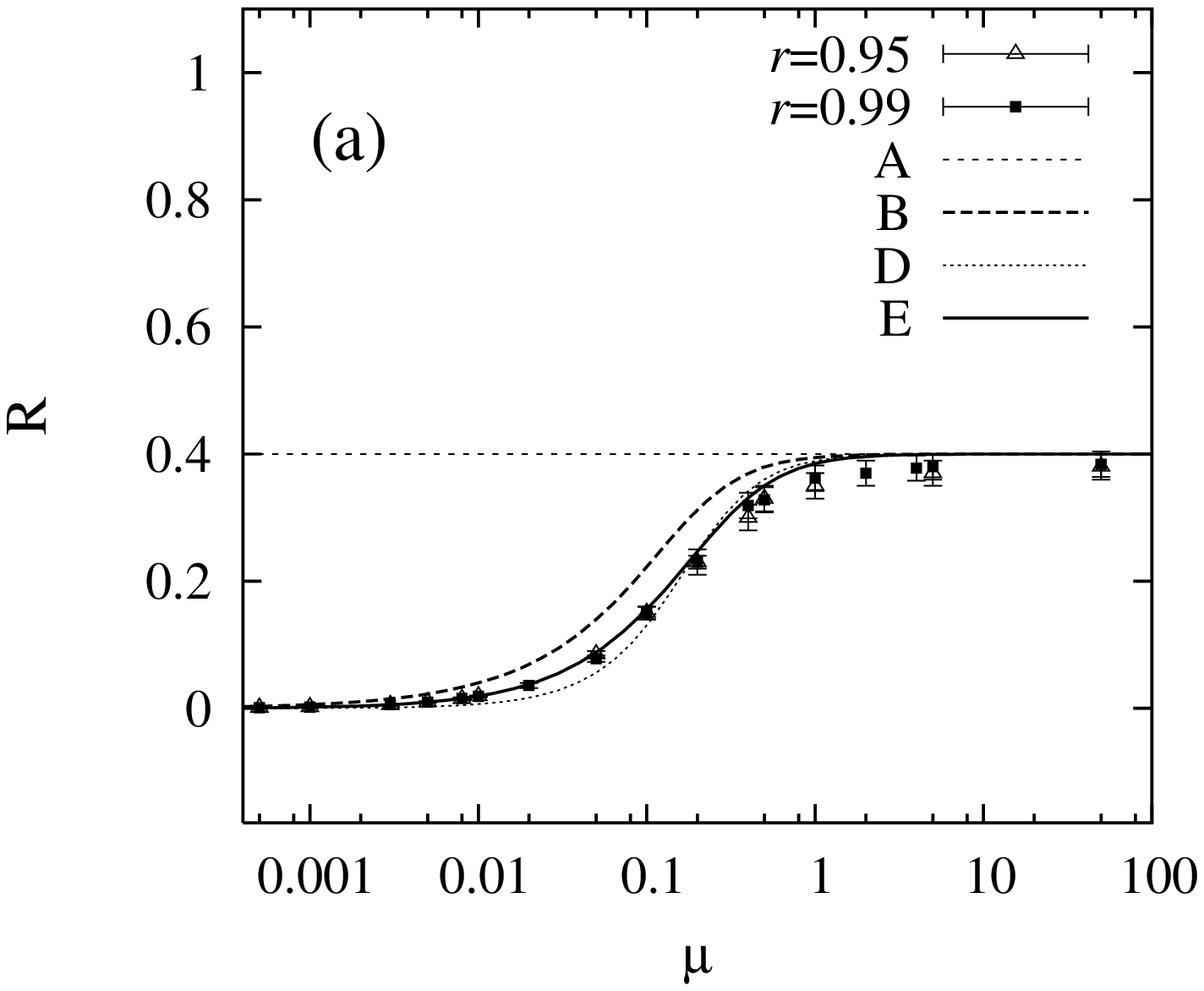,height=7.5cm,angle=-90}
  \epsfig{file=\picdirectory/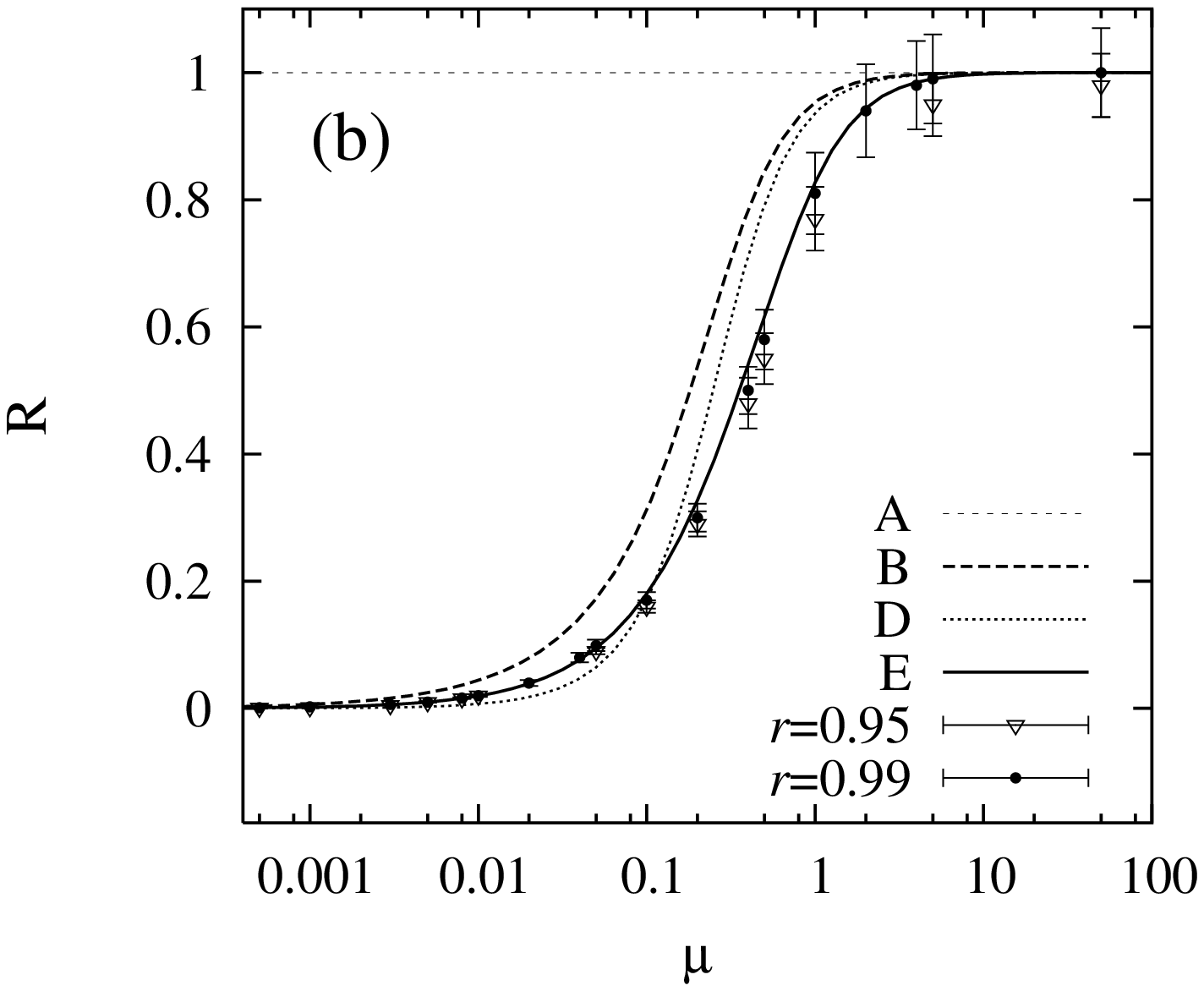,height=7.5cm,angle=-90}
\end{center}
\caption{
  $R$ plotted against $\mu$, for the same parameters as in Fig.\ 
  \protect{\ref{fig:Tmu}}.  The tangential restitution coefficients
  are again fixed to (a) $r_t^m=0.4$, and (b) $r_t^m=1.0$. }
\label{fig:Rmu}
\end{figure}

In Fig.\ \ref{fig:Rmu} we plot the ratio of rotational and
translational temperature in the same way as in Fig.\ 
\ref{fig:varyrtm}(c), but now, like in Fig.\ \ref{fig:Tmu}, we keep
the values $r_t^m=0.4$ (a) and $r_t^m=1.0$ (b) fixed and vary $\mu$.
Also here, we compare data for $r=0.99$ and $r=0.95$ in one plot.  For
the values of $r_t^m$ examined (see Fig.\ \ref{fig:Rmu}) one observes
a smooth transition of $R$ over about three orders of magnitude in
$\mu$, from the value $R=0$ (in the limit $\mu=0$) to the value
$R=r_t^m$ (in the limit $\mu \to \infty$). Note that the observation
$R=r_t^m$ is coincidence, since the correct asymptotic result for
large $\mu$ is $R=2(1+r_t^m)/(9-5r_t^m)$.  Again, the agreement
between simulations and model E is impressive.

\begin{figure}[htb]
\begin{center}
\epsfig{file=\picdirectory/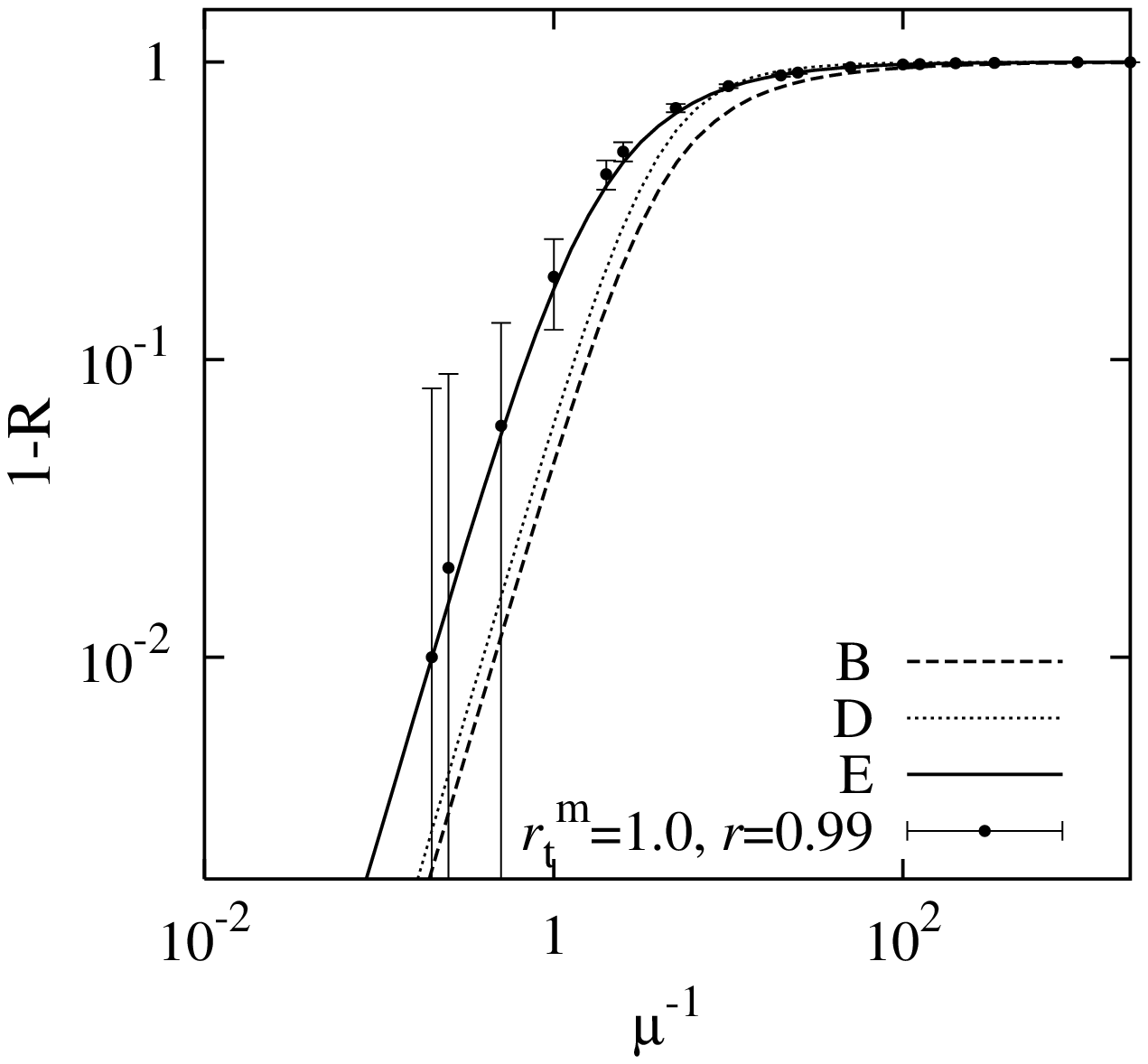,height=7.5cm,angle=-90}
\end{center}
\caption{Deviation from equipartition, $1-R$,
  plotted against the inverse friction coefficient, $\mu^{-1}$, for
  simulations from Fig.\ \protect{\ref{fig:Tmu}}(b). Note the
  double-logarithmic scale of this plot.  }
\label{fig:largemu}
\end{figure}

All models agree qualitatively in the large $\mu$-limit, even though
the quantitative agreement with simulations is again best caught by
model E, as can be seen in Fig.\ \ref{fig:largemu}.

\begin{figure}[htb]
\begin{center}
\epsfig{file=\picdirectory/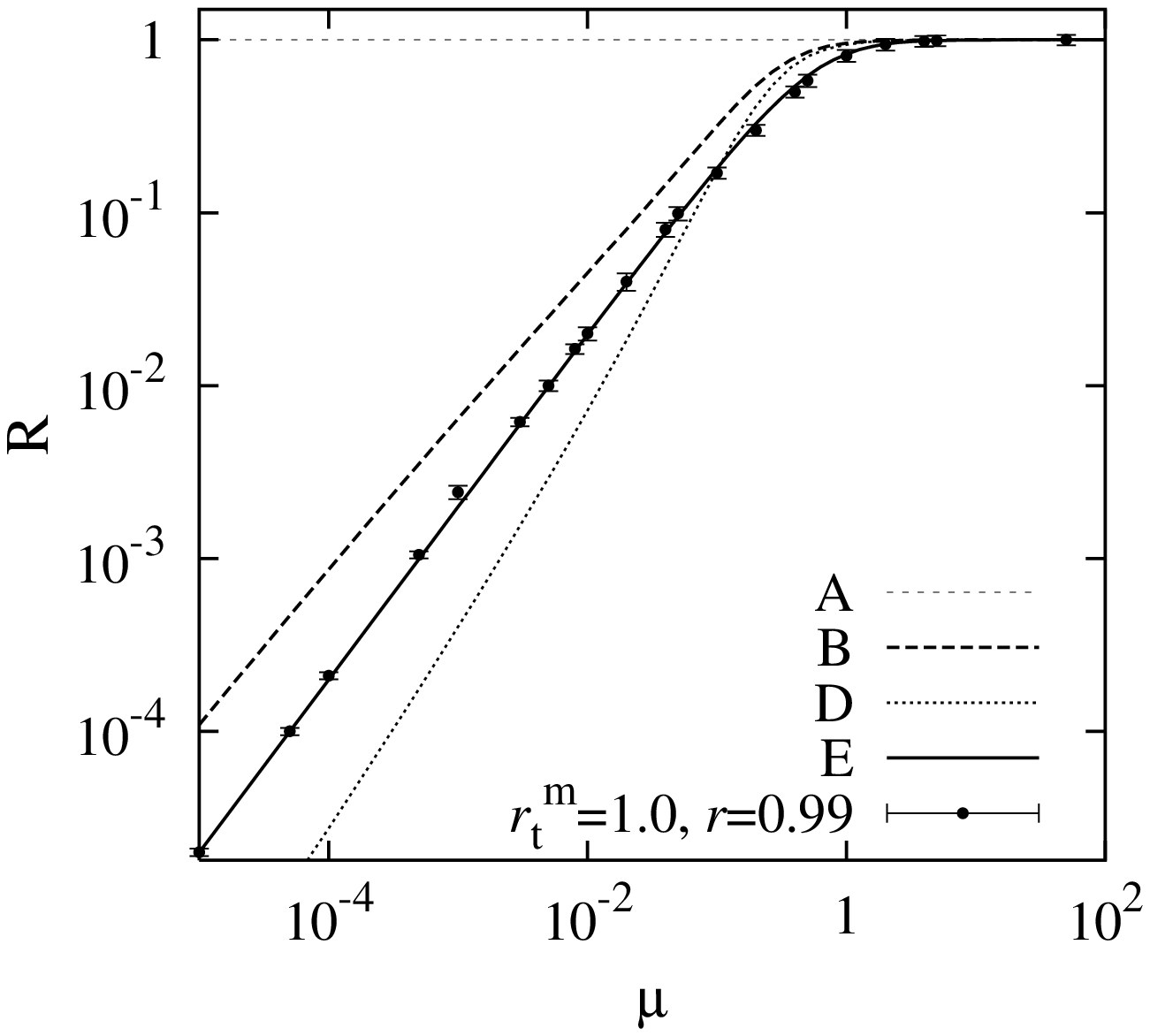,height=7.5cm,angle=-90}
\end{center}
\caption{Ratio of rotational and translational temperature,
  $R$, plotted against $\mu$, for some simulations from Fig.\ 
  \protect{\ref{fig:Tmu}}(b).  Note the double-logarithmic scale of
  this plot.  }
\label{fig:Rmus}
\end{figure}

The remaining question is the asymptotic behavior for very small
$\mu$, as can be viewed in Fig.\ \ref{fig:Rmus}, and as discussed
theoretically in subsection\ \ref{sec:steadystate_analytical}.  The
quantitative behavior of $R$ for small $\mu$ is tested by a power law
fit of the numerical values, according to an expression $R= b
\mu^{\alpha}$.  The fit gives $\alpha=1.00(4)$, for $r=0.99$,
$r_t^m=0.4,1.0$ and $\alpha=0.99(4)$, for $r=0.95$, $r_t^m=0.4,1.0$.
Thus the asymptotic behavior is proportional to $\mu$, in excellent
qualitative {\em and} quantitative agreement with the prediction of
model E.

\section{Approach to steady state}
\label{sec:approach.to.steady.state}

\subsection{Close to steady state}

Provided the system is sufficiently close to steady state, we can
linearize the set of Eqs.\ (\ref{eq:einfacheDGLs}) around $T_{\rm
  tr}^{\rm stat}$ and $T_{\rm rot}^{\rm stat}$. This is particularly
simple for models A, B, and D, where the coefficients in the
differential equation do not depend on $R$ and hence can be solved
analytically for the stationary state. We set $T_{\rm tr}(t)=T_{\rm
  tr}^{\rm stat}(1+\delta T_{\rm tr}(t))$ and $T_{\rm rot}(t)=T_{\rm
  rot}^{\rm stat}(1+\delta T_{\rm rot}(t))$ and obtain the linearized
dynamic equations
\begin{eqnarray}
\frac{d}{dt} \delta T_{\rm tr} & = & G T_{\rm tr}^{\rm stat}\left\{ 
      \left ( \frac{3}{2}A +\frac{B B'}{2C} \right ) \delta T_{\rm tr}
     +\frac{B B'}{C} \delta T_{\rm rot} \right\}\,,\nonumber\\
\frac{d}{dt} \delta T_{\rm rot} & = & 2 G C T_{\rm tr}^{\rm stat}\left\{ 
 \delta T_{\rm tr}- \delta T_{\rm rot} \right\} \,.
\end{eqnarray}
This set of linear equations is easily solved to yield two relaxation
rates $\lambda_1$ and $\lambda_2$.  In a stable stationary state they
must be positive and they are. We present here only results for the
simplest model (A) and postpone the general discussion to the next
paragraph, where the full dynamic evolution towards steady state will
be examined.

\begin{figure}[htb]
\begin{center}
\epsfig{file=\picdirectory/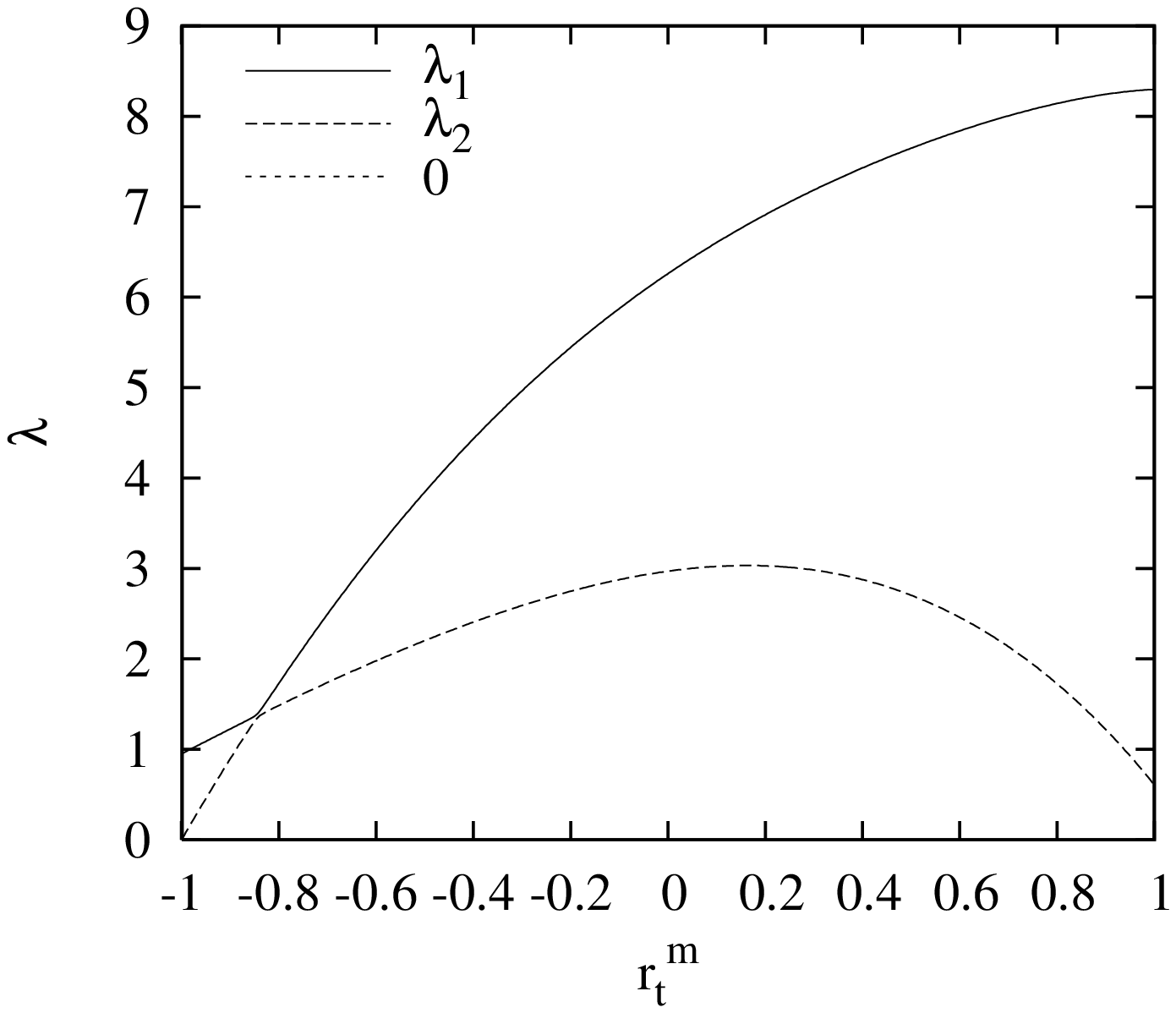,height=7.5cm,angle=-90}
\end{center}
\caption{Relaxation rates $\lambda_{1,2}$, close to steady state
         for $r=0.95$ as a function of $r_t^m$.}
\label{fig:lambda_12}
\end{figure}

In Fig.\ \ref{fig:lambda_12}, we plot the two relaxation rates as a
function of $r_t^m$ for a fixed value of $r=0.95$. In the limit of
smooth spheres one of the rates vanishes because the rotational energy
is conserved in that limit. For $r_t^m \sim - 0.84$ the two rates are
equal and for increasing $r_t^m$ the difference between the two rates
increases monotonically with $r_t^m$, such that for perfectly rough
spheres the larger rate is about fourteen times the smaller one. Such
a pronounced separation of time scales is familiar from the cooling
dynamics of the same model, see \cite{herbst00}. There it was shown
that the ratio of translational to rotational energy, $R$, relaxes
fast to its stationary value, whereas both the translational as well
as the rotational energy decay on the same, much longer time scale.
This point will be discussed in a more general setting (model E and
relaxation from an arbitrary initial condition) in the subsequent
paragraph.

\subsection{Full Dynamic Evolution}

\begin{figure}[htb]
\begin{center}
\epsfig{file=\picdirectory/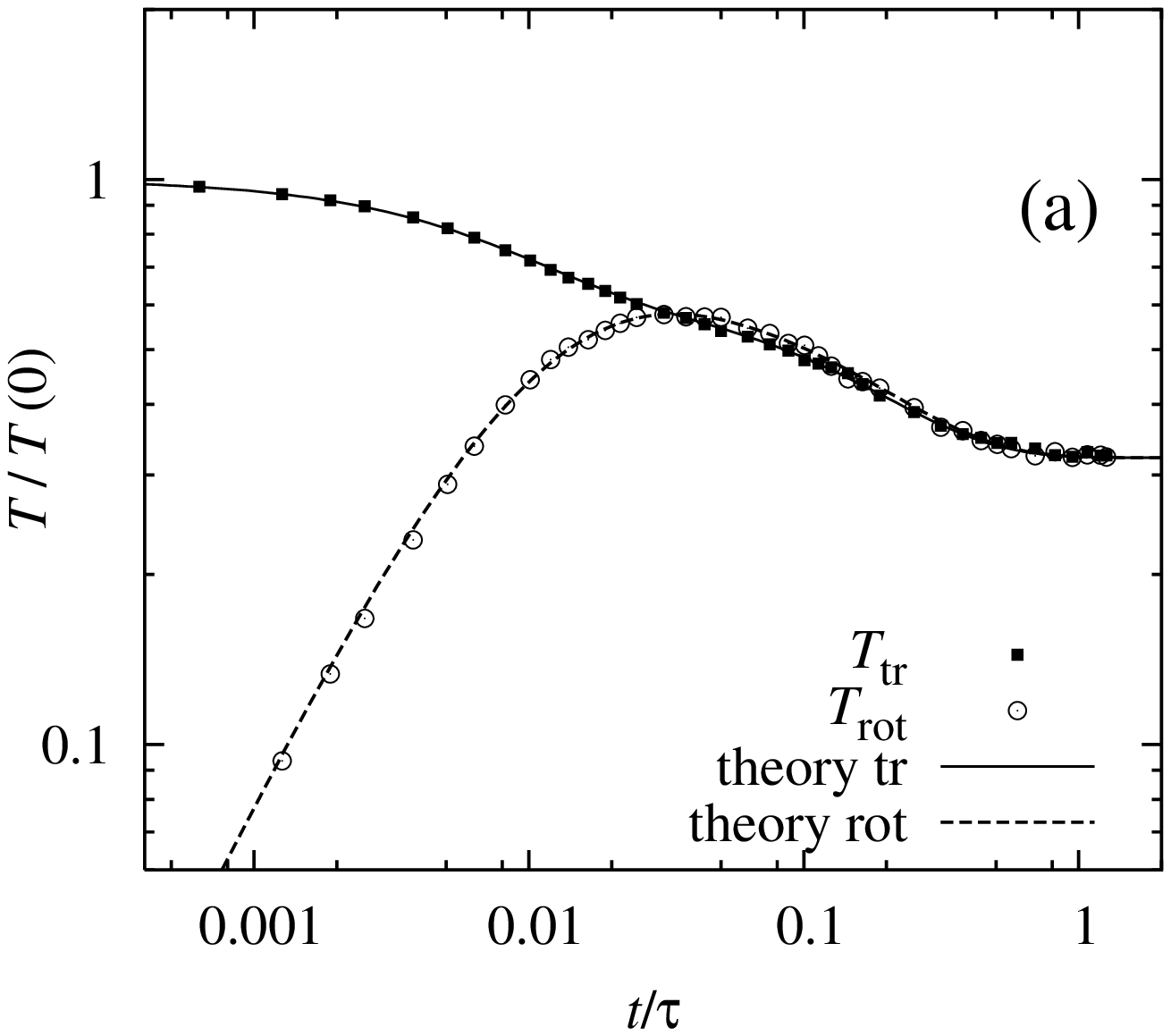,height=7.5cm,angle=-90} ~\\
\epsfig{file=\picdirectory/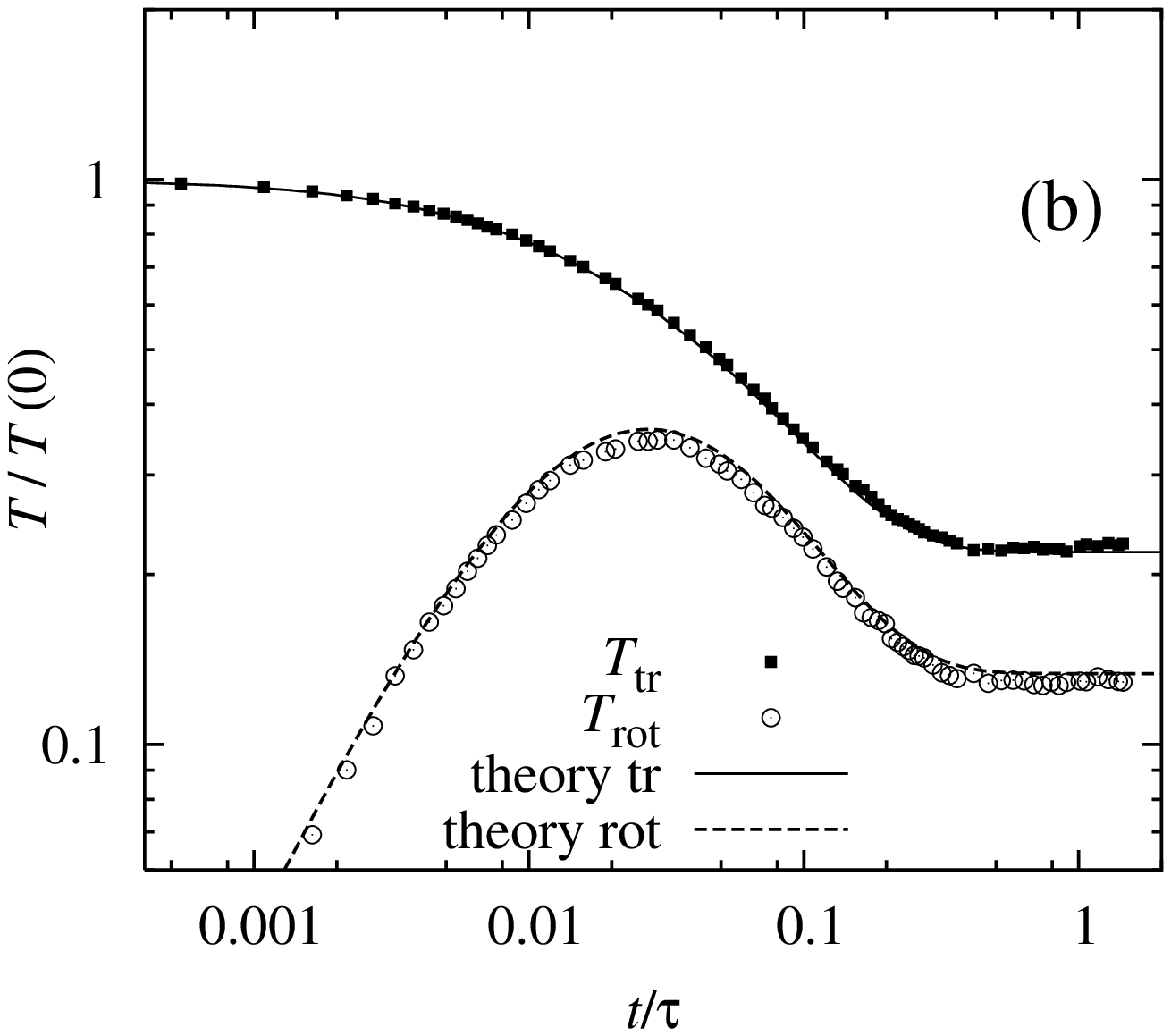,height=7.5cm,angle=-90} ~\\
\end{center}
\caption{Evolution of temperatures with rescaled time, 
  with $\tau^{-1}=(1/2) G T_{\rm tr}(0)^{1/2}$, for simulations with
  $N=11025$, $\nu=0.0866$, $r=0.95$, $r_t^m=1.0$, and (a)
  $\mu=\infty$, (b) $\mu=0.5$.}
\label{fig:T_time}
\end{figure}

In Fig.\ \ref{fig:T_time}, the full dynamic evolution of the
translational and rotational temperatures with time is shown for two
simulations with $N=11025$, $\nu=0.0866$, $r=0.95$, $r_t^m=1.0$, and
different values for the coefficient of friction. In both situations,
the agreement between simulations and the numerical solution for the
full MF theory, model E, is good -- not only concerning the limiting
values and the asymptotes, but also the time dependence during the two
regimes (i) equilibration between $T_{\rm tr}$ and $T_{\rm rot}$, and
(ii) approach to final steady state.

\section{Summary and Discussion}
\label{sec:conclusions}

In summary, a dynamic MF theory for the full time evolution of the
translational and rotational temperatures of a homogeneously driven
two-dimensional granular gas has been presented.  Particle collisions
were modeled using the Walton model \cite{walton86}, i.e.\ with normal
dissipation, tangential restitution (sticking) and Coulomb friction
(sliding). The Walton model can be formulated in terms of a
coefficient of tangential restitution, which {\it depends} on the
impact angle $\gamma$. Using a Pseudo-Liouville operator we have
computed the distribution of impact angles as well as the mean field
dynamics and steady state values of the translational and rotational
temperatures.

In addition to the complete mean field theory of the Walton model
(``model E''), we discussed three levels of approximation in order to
simplify the differential equations of the time evolution.  The
crudest approximations including Coulomb friction (``model B'' and
``model C'') assume that an effective constant tangential restitution
exists and can be computed by averaging over the angular distribution
of impact angles.  For model C this averaged coefficient depends on
the current values of the translational and rotational temperatures
and thus on time.  Even simpler is model B where the rotational
contribution to the impact angle is neglected, leading to a
coefficient of tangential restitution that only depends on global
system parameters.  The closest approximation (``model D'') to the
full mean field theory (``model E'') keeps the dependence of
$r_t(\gamma)$ on the impact angle $\gamma$ but, like for model B, the
contribution of the rotation of the particles to the impact angle is
neglected.

The predictions of the increasingly refined models of frictional
dissipation as well as the full MF theory have been compared to
simulations of a randomly driven mono-layer of spheres using an Event
Driven algorithm.  Emphasis has been put on the stationary state which
is characterized by two temperatures, $T_{\rm tr}$ and $T_{\rm rot}$,
one for the translational and one for the rotational degrees of
freedom.  Guided by the MF approach we discovered a rich phenomenology
like a non-trivial dependence of the stationary state temperatures on
the model parameters.  For example, the translational temperature is
non-monotonic as a function of maximal tangential restitution $r_t^m$
and also non-monotonic as a function of Coulomb friction $\mu$,
provided $r_t^m$ is sufficiently large.

All models predict steady state values of the translational and
rotational temperatures, which are considerably improved as compared
to the model without friction (``model A''), which assumes constant
tangential restitution (see Figs.\ \ref{fig:Tmu} and \ref{fig:Rmu}).
All approximations A-E agree in the limit of large friction, where the
tangential restitution becomes independent of the impact angle (see
Fig.\ \ref{fig:r_t}).  Qualitative agreement between models B-D and
simulations is achieved also for intermediate values of $\mu$. However
in the limit $\mu \to 0$ all approximations break down and only the
complete mean field solution (``model E'') is in agreement with the
simulations (see Fig.\ \ref{fig:Rmus}). In particular model E predicts
the linear dependence of the ratio of temperatures, $R=T_{\rm
  rot}/T_{\rm tr}$, on the friction coefficient $\mu$ that is observed
in the simulations and was used in Ref.\ \cite{jenkins02} to derive an
approximate kinetic theory of frictional particles.

Sticking contacts become more important relative to sliding contacts
for fixed $\mu$ and decreasing $r_t^m$. In this regime models B and D
seem reasonable, but lead to poor quantitative agreement as $r_t^m$
approaches $1$. The full mean field theory (``model E'') leads to
reasonable agreement for all values of $r_t^m$. For weak dissipation,
$r \to 1$, the agreement is very good -- for stronger dissipation, we
relate the deviations to the failure of both the homogeneity
assumption and the molecular chaos assumption made.

Linearizing the dynamic MF equations around the steady state leads to
an eigenvalue problem with two relaxation rates, one of them being
related to the equilibration between the translational and the
rotational degrees of freedom, while the other one controls the
approach of the system to its steady state.  For strong coupling, the
former process is much faster, so that there is a clear separation of
time scales, which has been discussed already for a freely cooling
system in the absence of driving.

In conclusion, realistic Coulomb friction turned out to be a subtle
problem as only the full mean field theory of the Walton model
predicts the effects of friction for all values of $\mu$ and $r_t^m$.
All simplifications are both qualitatively and quantitatively wrong in
some parameter range. Our studies can easily be extended to three
dimensional systems or more complex ones, like e.g. a polydisperse
mixture of frictional particles with different material properties.
Other driving mechanisms could be employed as well.

\section*{Acknowledgements}
This work has been supported by the European network project
FMRXCT980183 (RC) and the DFG (Deutsche Forschungsgemeinschaft)
through SFB 602 (OH,AZ), Grant No.Zi209/6-1 (AZ).  SL was also
supported by the DFG, Grant No.Lu450/9-1, and by FOM (Stichting
Fundamenteel Onderzoek der Materie, The Netherlands) as financially
supported by NWO (Nederlandse Organisatie voor Wetenschappelijk
Onderzoek).

\bibliographystyle{apsrev}

\begin{appendix}

\section{Derivation of the differential equations for model E}
\label{appendix:modelE}

The details of the derivation of the coefficients in Eqs.\ 
(\ref{eq:const3dnew}) and (\ref{eq:const3dnew2}) for model E in Sec.\ 
\ref{sec:full.version} will be shown.  The calculations are performed
using a Pseudo-Liouville operator formalism
\cite{ernst69,luding98d,herbst00}.  They are very similar to the ones
in three dimensions \cite{herbst00}.  First, we briefly recall the
Pseudo-Liouville operator formalism.

Let the vectors of position, translational and rotational velocity of
a particle $k$, in a two-dimensional plane ($x$,$y$) with only
vertical spin ($z$), be defined as
$\boldsymbol{r}_k=\left(r_{k,x},r_{k,y},0\right)$,
$\boldsymbol{v}_k=\left(v_{k,x},v_{k,y},0\right)$, and
$\boldsymbol{\omega}_k=\left(0,0,\omega_{k}\right)$ ~.

The time evolution of a dynamic variable $A(t)$ that depends on time
only through the positions and velocities of $N$ particles, can be
determined by means of a pseudo-Liouville operator ${\cal L}_{+}$ for
$t > 0$
\begin{equation}
\begin{split}
  A(t) 
  =\exp(i{\cal L}_{+}t)A(0) ~.
\end{split}
\end{equation}
The pseudo-Liouville operator ${\cal L}_{+}$ consists of three parts
${\cal L}_{+}={\cal L}_0 + {\cal L}^{'}_{+} + {\cal L}^{H}_{+}$. The
last part, ${\cal L}^{H}_{+}$, describes the homogeneous driving, the
first one, ${\cal L}_0$, describes the free streaming of particles
\begin{equation}
  {\cal L}_0 = -i \sum_{k=1}^{N} \boldsymbol{v}_{k} \cdot
  {\nabla}_{\boldsymbol{\scriptstyle r}_{k}},
\end{equation}
and the second one, ${\cal L}^{'}_{+}=\frac{1}{2}\sum_{l\ne
  k}T_{+}^{kl}$ describes hard-core collisions of two particles
\begin{equation}\label{stossit}
 T_{+}^{kl} = i(\boldsymbol{v}_{kl}\cdot\hat{\boldsymbol{r}}_{kl})
 \Theta(-\boldsymbol{v}_{kl}\cdot\hat{\boldsymbol{r}}_{kl})
 \delta(|\boldsymbol{r}_{kl}|-2a)(b_{+}^{kl}-1).
\end{equation}
The operator $b_{+}^{kl}$ replaces the linear and angular momenta of
the two particles $k$ and $l$ before collision by the corresponding
ones after collision, according to 
Eqs.\ (\ref{eq:linearmomentumchange}) and (\ref{eq:tangentialmomentumchange}).
$\Theta(x)$ is the Heaviside step--function, and we have introduced
the notation
$\boldsymbol{{r}}_{kl}=\boldsymbol{r}_{k}-\boldsymbol{r}_{l}$ and
$\hat{\boldsymbol{r}}_{kl}=\boldsymbol{r}_{kl}/|\boldsymbol{r}_{kl}|$.
Equation (\ref{stossit}) has the following interpretation: The factor
$\boldsymbol{v}_{kl}\cdot\hat{\boldsymbol{r}}_{kl}$ gives the flux of
incoming particles, while the $\Theta$- and $\delta$-functions specify the
conditions for a collision to take place. A collision between
particles $k$ and $l$ happens only if the two particles are
approaching each other which is ensured by
$\Theta(-\boldsymbol{v}_{kl}\cdot\hat{\boldsymbol{r}}_{kl})$. At the
instant of a collision the distance between the two particles has to
vanish when two particles touch, which is expressed by 
$\delta(|\boldsymbol{r}_{kl}|-2a)$. Finally,
$(b_{+}^{kl}-1)$ generates the change of linear and angular momenta
according to Eqs.\ 
(\ref{eq:linearmomentumchange}) and (\ref{eq:tangentialmomentumchange}).

The ensemble average, $\langle ... \rangle_t$, of a dynamic variable,
$A$, is defined by
 \begin{equation}\label{ensav1}
 \begin{split}
   \langle A \rangle _t & = \int d\Gamma \rho (0) A(t)=\int d\Gamma
   \rho (t) A(0)\\& = \int \prod_{k=1}^{N}(d^2{r_{k}} d^2{v_{k}}
   d{\omega_{k}}) ~ \rho (t) A(0) ~.
 \end{split}
 \end{equation}
 Here $\rho(t)=\exp {(-i{\cal L}_+^{\dagger}t)}\,\rho(0)$ is the
 $N$-particle distribution function, whose time development is
 governed by the adjoint ${\cal L}_+^{\dagger}$ of the time evolution
 operator ${\cal L}_+$.  Differentiating equation (\ref{ensav1}) with
 respect to time yields
 \begin{equation}\label{ensav2}
 \begin{split}
   \frac{d}{dt} \langle A \rangle _t &= \int d\Gamma \rho (0)
   \frac{d}{dt}
   A(t)=\int d\Gamma \rho (0) i{\cal L}_+ A(t)\\
   &=\int d\Gamma \rho (0)
   \exp{(i {\cal L}_+ t)} i{\cal L}_+ A(0)\\
   &=\int d\Gamma \rho (t) i{\cal L}_+ A(0) = \langle i{\cal L}_+ A
   \rangle _t ~.
 \end{split}
 \end{equation}
 
 The observables of interest are the averaged energies per particle,
 or, more specifically, the granular temperatures for the
 two-dimensional system
 \begin{align}\nonumber
   T_{ {\rm tr}} := \frac{ E_{ {\rm tr}} }{N}
   =& \frac{1}{N}\sum_{k=1}^{N}\frac{m}{2}|\boldsymbol{v}_{k}|^2\\
   \frac{1}{2} T_{ {\rm rot}} := E_{ {\rm rot}} 
   =& \frac{1}{N}\sum_{k=1}^{N}\frac{I}{2}|\boldsymbol{\omega}_{k}|^2
 \end{align}
 and the total kinetic energy $E=E_{ {\rm tr}}+E_{ {\rm rot}}$.  To
 make the temperatures dimensionless we may choose to measure mass in
 units of the particle mass, and velocities in units of the driving
 velocity $v_{0}$ defined in Eq.\ (\ref{eq:hom.driving}).
 
 Assuming a homogeneous density distribution and Gaussian velocity
 distributions the $N$-particle distribution function is given by

 \begin{equation}\label{hcs}
   \rho(t) \propto \prod_{k<l} \Theta(|\boldsymbol{r}_{kl}|-2a) 
                   \exp{\left\{-\left(\frac{E_{{\rm tr}}}
                       {T_{{\rm tr}}(t)}+\frac{E_{{\rm rot}}}
                       {T_{{\rm rot}}(t)}\right)\right\}} ~,
 \end{equation}
 where the product of Heaviside functions accounts for the excluded
 volume.  Hence we get two coupled differential equations for the time
 evolution of the translational and rotational energies
 \begin{align}\nonumber
   \frac{d}{dt} T_{{\rm tr}}(t) =& \frac{d}{dt}\langle E_{ {\rm
       tr}}\rangle_{t} = \langle i{\cal L}_{+}^{}E_{ {\rm
       tr}}\rangle_{t} = H_{\rm dr} + \langle i{\cal L}_{+}^{'}E_{
     {\rm tr}}\rangle_{t}\\\label{appendix.A.1} \frac{1}{2}
   \frac{d}{dt} T_{{\rm rot}}(t) =& \frac{d}{dt}\langle E_{ {\rm
       rot}}\rangle_{t} = \langle i{\cal L}_{+}^{}E_{ {\rm
       rot}}\rangle_{t} = \langle i{\cal L}_{+}^{'}E_{ {\rm
       rot}}\rangle_{t} ~.
 \end{align}
 The averages on the right hand sides can be calculated as follows.
 These calculations are almost identical for the translational and
 rotational energies, so we will show in detail the time derivative of
 $T_{{\rm tr}}(t)$ only.
\begin{equation}\lab{appendix.A.2}
\begin{split}
  \langle i{\cal L}^{'}_{+}E_{ {\rm tr}}\rangle_{t} &= \langle
  \frac{1}{2}\sum\limits_{k\ne l}\limits^{} iT^{kl}_{+}E_{ {\rm
      tr}}\rangle_{t}\\&=\frac{1}{2N}
  \sum\limits_{k=1}\limits^{N}\sum\limits_{\underset{l\ne
      k}{l=1}}\limits^{N}
  \frac{m}{2}\int\prod\limits_{j=1}\limits^{N}{d^2r_jd^2v_jd\omega_j}\rho(\boldsymbol{r},\boldsymbol{v},\boldsymbol{\omega})\times
  \\ &\quad\quad iT^{kl}_{+}
  (|\boldsymbol{v}_k|^2+|\boldsymbol{v}_l|^2)\\
\end{split}
\end{equation}
We have used that the binary collision operator $iT^{kl}_{+}$ yields
zero acting on any variable other than the ones of the two particles
involved in the collision.  Defining
\begin{equation}\lab{appendix.A.2x1}
\begin{split}
  d\Gamma :&= \prod\limits_{j=1}\limits^{N}{d^2r_jd^2v_jd\omega_j} \prod
  \limits_{l \ne j}\limits^{} \Theta(|\boldsymbol{r}_{jl}|-{2{a}})\times \\
  &\quad\quad
  \exp{\left(-\sum\limits_{k=1}\limits^{N} {\frac{m}{2T_{\rm
          tr}(t)}|\boldsymbol{v}_k|^2}-\sum\limits_{k=1}\limits^{N}{\frac{I}{2T_{\rm
          rot}(t)}|\boldsymbol{\omega}_k|^2}\right)}
\end{split}
\end{equation}
and using the definition of $iT^{12}_{+}$ we can write
\begin{equation}\lab{appendix.A.3}
\begin{split}
  \langle i{\cal L}^{'}_{+}E_{ {\rm tr}}\rangle_{t} &= -\frac{N-1} {2\int d\Gamma}
    \int d\Gamma~
    (\hat{\boldsymbol{r}}_{12}\cdot{\boldsymbol{v}_{12}})~\delta{(|\boldsymbol{r}_{12}|-{2{a}})}\times\\&\quad\quad \Theta{(-\hat{\boldsymbol{r}}_{12}\cdot\hat{\boldsymbol{v}}_{12})}(b^{12}_{+}-1)
    \frac{m}{2} (|\boldsymbol{v}_1|^2+|\boldsymbol{v}_2|^2)
\end{split} \nonumber
\end{equation}
The change of energy $\Delta E_{\rm tr}:= \frac{m}{2}
(b^{12}_{+}-1)(|\boldsymbol{v}_1|^2+|\boldsymbol{v}_2|^2)$ that
results from a collision of particle 1 and 2 depends only on the phase
space variables of particle 1 and 2.  Since we assume spatial
homogeneity this change of energy can only depend on the relative
distance vector
$\boldsymbol{r}_{12}:=\boldsymbol{r}_{1}-\boldsymbol{r}_{2}$ as well
as the relative translational and rotational velocities
$\boldsymbol{v}_{12}:=\boldsymbol{v}_{1}-\boldsymbol{v}_{2}$ and
$\boldsymbol{\omega}_{12}:=\boldsymbol{\omega}_{1}+\boldsymbol{\omega}_{2}$.
Further, we assume instantaneous collisions.  Therefore the change of
energy can only depend on the direction of the distance vector
$\hat{\boldsymbol{r}}_{12}=\frac{\boldsymbol{r}_{1}-\boldsymbol{r}_{2}}{|\boldsymbol{r}_{1}-\boldsymbol{r}_{2}|}$.
Now we can perform the integrations over 
those particles that are not involved in the collision.  The integrals
over $d^2\boldsymbol{v}_3 \hdots d^2\boldsymbol{v}_N$ and
$d^2\boldsymbol{\omega}_3 \hdots d^2\boldsymbol{\omega}_N$ are simple
Gaussians.  To integrate over $d^2\boldsymbol{r}_3 \hdots
d^2\boldsymbol{r}_N$ we introduce two more two-dimensional integrals
$${\int
    d^2\boldsymbol{R}_1d^2\boldsymbol{R}_2
    \delta^2{(\boldsymbol{R}_1-\boldsymbol{r}_1)}
    \delta^2{(\boldsymbol{R}_2-\boldsymbol{r}_2)}} ~, $$  
over two-dimensional $\delta$ functions, 
$\delta^2(\boldsymbol{r}):= \delta{(r_x)}\delta{(r_y)}$,
Using the definition of the pair correlation function
\begin{equation}\lab{appendix.A.5}
\begin{split}
  &\frac{g(|\boldsymbol{R}_1-\boldsymbol{R}_2|)}{V^2}:=\\ &\quad \frac{\int
    \prod\limits_{j=1}\limits^{N}{d^2r_j} \prod \limits_{l\ne
      j}\limits^{}
    \Theta(|\boldsymbol{r}_{jl}|-{2{a}})\delta^2{(\boldsymbol{R}_1-\boldsymbol{r}_1)}\delta^2{(\boldsymbol{R}_2-\boldsymbol{r}_2)}}{\int
    \prod\limits^{N}\limits_{j=1}d^2r_j \prod \limits_{l\ne
      j}\limits^{} \Theta(|\boldsymbol{r}_{jl}|-{2{a}})} ~,\end{split} \nonumber
\end{equation}
where $V$ is the area of the system, we obtain
\begin{equation}\lab{appendix.A.6}
\begin{split}
  &\langle i{\cal L}^{'}_{+}E_{ {\rm tr}}\rangle_{t}=
  -\frac{N-1}{2V^2}\left(\frac{m}{2\pi T_{\rm
        tr}(t)}\right)^{2}\left(\frac{I}{2\pi T_{\rm
        rot}(t)}\right)^{}\\&\quad \int
  d^2{R_1}d^2{R_2}
  d^2{v_{1}}d^2{v_{2}}d^2\omega_{1}d\omega_{2}
  ~\ g(|\boldsymbol{R}_{12}|)\times \\&\quad \exp
  \left(\frac{-\frac{m}{2}(|\boldsymbol{v}_1|^2+|\boldsymbol{v}_2|^2)}{T_{\rm
        tr}(t)} -
    \frac{\frac{I}{2}(|\boldsymbol{\omega}_1|^2+|\boldsymbol{\omega}_2|^2)}{T_{\rm
        rot}(t)}\right) \times \\ & \quad
  (\hat{\boldsymbol{R}}_{12}\cdot{\boldsymbol{v}_{12}})~\delta{(|\boldsymbol{R}_{12}|-{2{a}})}\Theta{(-\hat{\boldsymbol{R}}_{12}\cdot\hat{\boldsymbol{v}}_{12})}
~\Delta E_{\rm tr}.\nonumber
\end{split}
\end{equation}
Since the change of energy $\Delta E_{\rm tr}$ depends only on
$\boldsymbol{R}_{12}:=\boldsymbol{R}_{1}-\boldsymbol{R}_{2}$,
$\boldsymbol{v}_{12}$, and $\boldsymbol{\omega}_{12}$, we introduce
the variables
\begin{align}\nonumber
  \boldsymbol{r_{}} &:= \boldsymbol{R}_{1}-\boldsymbol{R}_{2}, \quad
  \boldsymbol{v_{}} :=
  \frac{\boldsymbol{v}_{1}-\boldsymbol{v}_{2}}{\sqrt{2}}, \quad
  \boldsymbol{\omega_{}} :=
  \frac{\boldsymbol{\omega}_{1}+\boldsymbol{\omega}_{2}}{\sqrt{2}}\\\label{eq:firstsubstitution}
  \boldsymbol{R_{}} &:= \boldsymbol{R}_{1}
  , \quad \boldsymbol{V}_{} :=
  \frac{\boldsymbol{v}_{1}+\boldsymbol{v}_{2}}{\sqrt{2}}, \quad
  {\color{black}\boldsymbol{\Omega}_{}} :=\frac{\boldsymbol{\omega}_{1}-\boldsymbol{\omega}_{2}}{\sqrt{2}} ~.
\end{align}
The Jacobian of this transformation is $1$.  The expression to
integrate over is independent of $\boldsymbol{R_{}}$ such that
integration over $d^2{R_{}}$ yields the area $V$. We write
$\boldsymbol{r_{}}$ in polar coordinates $(r, \phi)$ and can integrate
over $dr$.  Then, we choose the coordinate system for integrations
over $d^2v$ such that the unit vector $\hat{\boldsymbol{r_{}}}$ points
in the $y$-direction.  That means we can replace
$\hat{\boldsymbol{r_{}}}$ by the unit vector in the $y$-direction
$\hat{\boldsymbol{e}}_{y}$ and integrate over $d\phi$ which simply
yields $2\pi$. For readability, we use now the unit vector
$\hat{\boldsymbol{n_{}}}$ instead of $\hat{\boldsymbol{e_{}}}_y$.  The
integrals over $d^2{V_{}}$ and $d\Omega_{}$ are Gaussians, so that we
obtain
\begin{equation}\lab{appendix.A.8}
\begin{split}
  \langle i{\cal L}^{'}_{+}E_{ {\rm tr}}\rangle_{t} &= -{2 \pi
    \sqrt{2}}~{{a}}~n_0~g({2{a}})\left(\frac{m}{2\pi T_{\rm
        tr}(t)}\right)^{}\left(\frac{I}{2\pi T_{\rm
        rot}(t)}\right)^{\frac{1}{2}}\times\\ &\quad \int dv_1 dv_2
  d\omega_{} \exp{\left(\frac{-\frac{m}{2}[v_1^2+v_2^2]}{T_{\rm
          tr}(t)}+\frac{-\frac{I}{2}|\omega_{}|^2}{T_{\rm
          rot}(t)}\right)}\times \\ &\quad
  (\hat{\boldsymbol{n}}\cdot{\boldsymbol{v_{}}})
  ~\Theta{(-\hat{\boldsymbol{n}}\cdot\hat{\boldsymbol{v}_{}})} ~\Delta
  E_{\rm tr} ~, \nonumber
\end{split}
\end{equation}
with the number density $n_0:={(N-1)}/{V}\approx{N}/{V}$.  

To solve the integrals above we need to take a look at the change of
energy
\begin{equation}\lab{appendix.A.10}
\begin{split}
  \frac{2}{m} \Delta E_{\rm tr}& := (b^{12}_{+}-1)
  (|\boldsymbol{v}_1|^2+|\boldsymbol{v}_2|^2) \\
  &= (|\boldsymbol{v}'_1|^2+|\boldsymbol{v}'_2|^2) -
  (|\boldsymbol{v}_1|^2+|\boldsymbol{v}_2|^2)
  \\
  &~\hspace*{-0.3cm} \overset{{(\ref{eq:linearmomentumchange}),
      (\ref{eq:tangentialmomentumchange})}}{=} 4 \eta (\eta-1)
  (|\boldsymbol{v}|^2-
  (\hat{\boldsymbol{n}}\cdot\boldsymbol{v})^2) \\
  & \quad\quad - (1-{r}^2) (\hat{\boldsymbol{n}}\cdot\boldsymbol{v})^2
  +({2{a}}
  \eta)^2 |\hat{\boldsymbol{n}} \times \boldsymbol{\omega}|^2 \\
  & \quad\quad + {4{a}} \eta(2\eta-1) ~(\hat{\boldsymbol{n}} \times
  \boldsymbol{\omega}) \cdot \boldsymbol{v}
\end{split}
\end{equation}
with $\eta$ and ${r_t}(\gamma)$ given by Eqs.\ (\ref{eq:eta}) and
(\ref{eq:betagamma.nurCoulomb}).  Keep in mind that $\boldsymbol{v}$
and $\boldsymbol{\omega}$ have been defined as
$\boldsymbol{v}:=\boldsymbol{v}_{12}/\sqrt{2}$ and
$\boldsymbol{\omega}:=\boldsymbol{\omega}_{12}/\sqrt{2}$.  The
difference in calculation for models D and E comes into play at this
step {\em [For model D, at this point of the calculation, we would
  express $\boldsymbol{v}$ in polar coordinates $(v, \gamma_{12})$ and
  insert $\eta=\eta(\gamma_{12})$ instead of $\eta=\eta(\gamma)$
  ``assuming'' that $\gamma \approx \gamma_{12}$.  That way all
  integrals become Gaussians and can easily be solved.  In particular,
  the integrals over the last term in Eq.\ (\ref{appendix.A.10})
  vanish.]}, however, we will now go on with model E.  To perform the
integrations over $\boldsymbol{v}$ and $\boldsymbol{\omega}$ we
substitute
\begin{align}\nonumber
  \boldsymbol{\omega}_{\perp} &:= \sqrt{2} {{}}
  ~(\hat{\boldsymbol{n}}\times\boldsymbol{\omega_{}}) =(\sqrt{2}\ {{}}
  \omega,0,0)\\\label{eq:secondsubstitution} \boldsymbol{g_{}} &:=
  \sqrt{2}\left(\boldsymbol{v_{}}+ a\ 
    (\hat{\boldsymbol{n}}\times\boldsymbol{\omega_{}})\right) ~,
\end{align}
thus introducing the relative velocity of the contact point
$\boldsymbol{g_{}}$ as defined in Eq.\ (\ref{eq:def.g}).  The vector
$\boldsymbol{\omega}_{\perp}$ points in the $x$-direction (due to our
choice of coordinates) and
$|\boldsymbol{\omega}_{\perp}|=|\boldsymbol{\omega}_{12}|$, so that
$\boldsymbol{v}_{\rm rot}=a\boldsymbol{\omega}_{\perp}$.  The Jacobian
of this transformation is ${2}^{-\frac{3}{2}}$.  In terms of these new
variables $\Delta E_{\rm tr}$ reads
\begin{equation}\lab{appendix.A.14}
\begin{split}
  \frac{2}{m}\Delta E_{\rm tr}&=
  2 \eta (\eta-1) \left(|\boldsymbol{g_{}}|^2-
  (\hat{\boldsymbol{n}}\cdot\boldsymbol{g_{}})^2\right)\\&\quad  -
  \frac{1-{r}^2}{2}
  (\hat{\boldsymbol{n}}\cdot\boldsymbol{g_{}})^2 + 2 {{a}} \eta~
  \boldsymbol{g}\cdot\boldsymbol{\omega}_{\perp} ~,
\end{split}
\end{equation}
and we get
\begin{equation}\lab{appendix.A.15}
\begin{split}
  &\langle i{\cal L}^{'}_{+}E_{ {\rm tr}}\rangle_{t} = - \frac{2}
  {\sqrt{\pi}} ~{{a}}~n_0~g({2{a}}) \left(\frac{m}{4T_{\rm
        tr}(t)}\right)^{}\left(\frac{I}{4T_{\rm
        rot}(t)}\right)^{\frac{1}{2}} \times \\
  &\quad \int d^2{g_{}}~d{\omega}_{\perp}~
  (\hat{\boldsymbol{n}}\cdot{\boldsymbol{g_{}}})
  ~\Theta{(-\hat{\boldsymbol{n}}\cdot\hat{\boldsymbol{g}_{}})} ~ \exp
  \left( -\frac{qma^2} {4 T_{\rm rot}(t)}
    |\boldsymbol{\omega}_{\perp}|^2
  \right ) \times \\
  &\quad \exp \left( -\frac{m \left[ |\boldsymbol{g_{}}|^2 -
        2{{a}}~\boldsymbol{g}\cdot\boldsymbol{\omega}_{\perp} +
        a^2|\boldsymbol{\omega}_{\perp}|^2 \right ] }{4 T_{\rm tr}(t)}
  \right ) ~\Delta E_{\rm tr} ~.
\end{split}
\end{equation}
Next, we express $\boldsymbol{g}$ in polar coordinates $(g, \gamma)$
where $\gamma$ is {\it not} the usual angle between $\boldsymbol{g}$
and $\hat{\boldsymbol{e}}_{x}$ but instead -- as needed for
incorporating Coulomb friction -- the angle between $\boldsymbol{g}$
and $\hat{\boldsymbol{n}}$ (i.e.\ the angle between $\boldsymbol{g}$
and $\hat{\boldsymbol{e}}_{y}$, i.e.,
$\boldsymbol{g}=(-g\sin{\gamma},g\cos{\gamma})$).  Expression
(\ref{appendix.A.14}) reads now
\begin{equation}\lab{appendix.A.18}
\begin{split}
  \frac{2}{m}\Delta E_{\rm tr}& = 2 \eta (\eta-1) g^2 \sin^2{\gamma} -
  \frac{1-{r}^2}{2} g^2 \cos^2{\gamma}\\&\quad\quad - {2{a}} \eta g 
  \omega_{\perp}
  \sin{\gamma} ~,
\end{split}
\end{equation}
(note that $\eta=\eta(\cot{\gamma})$ in the Coulomb friction case) and
\begin{equation}\lab{appendix.A.19}
\begin{split}
  & \langle i{\cal L}^{'}_{+}E_{ {\rm tr}}\rangle_{t} =-
  \frac{2}{\sqrt{\pi}} ~{{a}}~ n_0 ~g({2{a}}) \left(\frac{m}{4 T_{\rm
        tr}(t)}\right)^{}\left(\frac{I}{4 T_{\rm
        rot}(t)}\right)^{\frac{1}{2}} \times \\
  & \int\limits_{\frac{\pi}{2}}\limits^{\frac{3}{2}\pi}d\gamma
  \int\limits_{0}\limits^{\infty} dg
  \int\limits_{-\infty}\limits^{\infty} d\omega_{\perp}
  ~g^2\cos{\gamma}
  ~ \exp \left( -\frac{qma^2 \omega_{\perp}^2}{4T_{\rm rot}(t)}\right) \times \\
  & \exp \left(-\frac{m}{4T_{\rm tr}(t)}
    \left[g^2+{2{a}}g\omega_{\perp}\sin{\gamma}+{a}^2\omega_{\perp}^2
    \right] \right) ~\Delta E_{\rm tr} ~.\nonumber
\end{split}
\end{equation}
Now we define $\mathcal{A}:= [m{a}^2/4][1/T_{\rm tr}(t)+q/T_{\rm
  rot}(t)]$, $\mathcal{B}:= {m{a}}\sin{\gamma}/[4{\mathcal{A}}T_{\rm
  tr}(t)]$ and substitute $p:= \sqrt{\mathcal{A}}\ 
(\omega_{\perp}+\mathcal{B}g)$ for $\omega_{\perp}$.  This leads to
Gaussian integrals over $p$ and $g$.  Using $x^2:= 1+\frac{T_{\rm
    rot}(t)}{qT_{\rm tr}(t)}$, we obtain
\begin{equation}
\begin{split}\lab{appendix.A.27}
  \langle i{\cal L}^{'}_{+}E_{ {\rm tr}}\rangle_{t} &= - \frac{3}{2}
  \sqrt{\frac{\pi}{m}} ~{2{a}} ~n_0 ~g({2{a}})~ T^{\frac{3}{2}}_{\rm
    tr}(t) ~x^4 \times \\
  &\quad\quad \int\limits_{\frac{\pi}{2}}\limits^{\pi}d\gamma
  \frac{\cos{\gamma}}{(1+(x^2-1)\cos^2{\gamma})^{\frac{5}{2}}}\times \\
  &\quad\quad \left(4\eta \left[ \eta-\frac{1}{x^2}\right]
    \sin^2{\gamma} - (1-{r}^2) \cos^2{\gamma}\right) ~.
\end{split}
\end{equation}
Up to this point we have {\it not} specified, whether we are going to
use constant coefficients of restitution or Coulomb friction.  All
this is hidden in $\eta$ which is either a constant or a function of
${\gamma}$.  Since we are interested in the Coulomb friction case, we
use $\eta=\eta(\gamma)$.  We introduce the notation
$\eta=\frac{1+{r}}{2} \mu\min{\{|\cot{\gamma_0}|, |\cot{\gamma}|\}}
\equiv \min{\{\eta_0, \frac{1+{r}}{2} \mu |\cot{\gamma}|\}}$, and
obtain
\begin{equation}\lab{appendix.A.29}
\begin{split}
  \langle i{\cal L}^{'}_{+}E_{ {\rm tr}}\rangle_{t} &= - \frac{3}{2}
  \sqrt{\frac{\pi}{m}} ~{2{a}}~ n_0~ g({2{a}}) T^{\frac{3}{2}}_{\rm
    tr}(t) ~x^4 \left\{ \frac{2}{3}\frac{1-{r}^2}{x^4} \right. \\
  & \left. \quad\quad+ \int\limits_{\frac{\pi}{2}}\limits^{\gamma_0}
    d\gamma \frac{\cos{\gamma}}{(1+(x^2-1)\cos^2{\gamma})^{5/2}}
    \times \right. \\ & \left. \left(2 \mu \frac{1+{r}}{x^2}
      \sin{\gamma}\cos{\gamma} + \mu^2 (1+{r})^2 \cos^2{\gamma}\right)
  \right. \\ & \left.  + 4\eta_0\left(\eta_0-\frac{1}{x^2}\right)
    \int\limits_{\gamma_0}\limits^{\pi}d\gamma
    \frac{\cos{\gamma}\sin^2{\gamma}}{(1+(x^2-1)\cos^2{\gamma})^{5/2}}\right\}
  ~.
\end{split}
\end{equation}

After performing the last integration the result can be written in the
form
\begin{eqnarray}
\label{eq:fullsolution.tr}
  \frac{d}{dt}T_{ {\rm tr}}(t) 
 &=& H_{\rm dr} -G T_{\rm tr}^{3/2}
  \left\{{\color{green} A_r} \right. \\\nonumber  &~&{\color{red}\hspace*{-0.2cm}+}\left.{\color{red}\frac{\eta_0}{2}} \frac{{\color{red}1- \eta_0
      x^2}} {(1+x^2\cot^2{\gamma_0})^{3/2}}
{\color{dark-green}+ \frac{\eta_0}{2}}\frac{{\color{dark-green}x^2\cot^2{\gamma_0}}}
  {(1+x^2\cot^2{\gamma_0})^{3/2}} \right. \\ \nonumber
&~&{\color{blue-green}\hspace*{-0.2cm}-}\left.{\color{blue-green}\eta_0^2\tan^2{\gamma_0}}
  \left({\color{red-green}1}-\frac{{\color{red-green}1} {\color{red-green}+ \frac{3}{2}x^2\cot^2{\gamma_0}}} {(1+x^2 \cot^2{\gamma_0})^{3/2}}\right) \right\} ~,
\end{eqnarray}
~\\
where $G=8\sqrt{\frac{\pi}{m}}~{a}~n_0~g({2{a}})$, which is the same
as Eq.\ (\ref{eq:G}), and $A_r=\frac{1-r^2}{4}$. Similarly $T_{\rm
  rot}$ can be calculated using, instead of $\Delta E_{\rm tr}$, the
change of rotational energy at collision,
\begin{equation}\lab{appendix.A.35}
\begin{split}
  \frac{2}{I}\Delta E_{\rm rot} &:=(b^{12}_{+}-1)
  (|\boldsymbol{\omega}_1|^2+|\boldsymbol{\omega}_2|^2) \\&=
  (|\boldsymbol{\omega'}_1|^2+|\boldsymbol{\omega'}_2|^2) -
  (|\boldsymbol{\omega}_1|^2+|\boldsymbol{\omega}_2|^2)\\
  &~\hspace*{-0.31cm}
  \overset{{(\ref{eq:linearmomentumchange}),(\ref{eq:tangentialmomentumchange})}}{=}
  \frac{4\eta^2}{a^2 q^2} |\hat{\boldsymbol{n}_{}} \times
  \boldsymbol{v_{}}|^2 + 4 \frac{\eta}{q}
  \left(\frac{\eta}{q}-1\right) \left(|\boldsymbol{\omega_{}}|^2 -
    \left(\hat{\boldsymbol{n}_{}}\cdot\boldsymbol{\omega_{}}\right)^2\right)
  \\&\quad\quad- \frac{4 \eta}{aq} \left(2\frac{\eta}{q}-1\right)
  \left(\hat{\boldsymbol{n}_{}} \times \boldsymbol{v_{}}\right) \cdot
  \boldsymbol{\omega_{}} ~,
 \end{split}
\end{equation}
which can be reformulated as
\begin{equation}\lab{appendix.A.35a}
 \begin{split}
   \frac{2}{m}\Delta E_{\rm rot} &= \frac{2\eta^2}{q}
   \left(|\boldsymbol{g_{}}|^2 -
     (\hat{\boldsymbol{n}}\cdot\boldsymbol{g_{}})^2\right) - 2a \eta ~
   \boldsymbol{g_{}}\cdot \boldsymbol{\omega}_{\perp} ~,
 \end{split}
\end{equation}
using the notation introduced in Eqs.\ (\ref{eq:firstsubstitution})
and (\ref{eq:secondsubstitution}).  The calculation for the rotational
temperature is identical to the one for the translational temperature
just shown until Eq.\ (\ref{appendix.A.15}) into which we insert
$\Delta E_{\rm rot}$ from Eq.\ (\ref{appendix.A.35a}) instead of
$\Delta E_{\rm tr}$.  Performing the integrals yields
\begin{eqnarray}
\label{eq:fullsolution.rot}
\frac{1}{2}\frac{d}{dt}T_{ {\rm rot}}(t)
&=&  G T_{\rm tr}^{3/2}
   \left\{ 
     \frac{{1}}
 {({1}+{x^2\cot^2{\gamma_0}})^{{3/2}}} ~\times
\right. \\ \nonumber
&~&\hspace*{-0.3cm}\left.
     \left({\color{red}\frac{\eta_0^2x^2}{2q}}~{\color{lila}-(x^2-1)\frac{\eta_0}{2}}({\color{red}1}+{\color{blue}x^2\cot^2{\gamma_0}})\right)
   \right. \\\nonumber  &~&\hspace*{-0.3cm}\left. 
{\color{blue-green}+\frac{\eta_0^2\tan^2{\gamma_0}}{q}}\left({\color{red-green}1}
  -\frac{{\color{red-green}1} {\color{red-green}+ \frac{3}{2}x^2\cot^2{\gamma_0}}} {(1+x^2
      \cot^2{\gamma_0})^{3/2}}\right)\right\}
\end{eqnarray}

~\\Finally, from Eqs.\ (\ref{eq:fullsolution.tr}) and
(\ref{eq:fullsolution.rot}) the conversant reader may reproduce the
transformation to the more convenient coefficients in 
Eqs.\ (\ref{eq:const3dnew}).

For comparison, we quote the equivalent results in three dimensions
\cite{herbst00}:
\begin{equation}\lab{appendix.A.32a}
\begin{split}
  \frac{3}{2}\frac{d}{dt} T_{{\rm tr}}(t) =~& H_{\rm dr} - G_{3D}
  T^{{3}/{2}}_{\rm tr} \left\{ {\color{green} A_r} \ {\color{red}+}\ 
    {\color{red}\frac{\eta_0}{2}}\frac{{\color{red}1- \eta_0 x^2}}
    {1+x^2\cot^2{\gamma_0}}
  \right. \\
  &~\hspace*{-0.2cm} \left. + \frac{\eta_0}{2}
    \left(\frac{\arctan{\left(x\cot{\gamma_0}\right)}}{x\cot{\gamma_0}}
      - \frac{1}{1+x^2\cot^2{\gamma_0}}\right) \right\},
\end{split}
\end{equation}
and
\begin{equation}\lab{appendix.A.43a}
\begin{split}
  \frac{3}{2}\frac{d}{dt} T_{{\rm rot}}(t) =~ &G_{3D} T^{{3}/{2}}_{\rm
    tr}
  \left\{{\color{red}\eta_0}\frac{{\color{red}1-\left(1-\frac{\eta_0}{q}\right)x^2}}{1+x^2
      \cot^2{\gamma_0}} \right. \\
  &~\hspace*{-1.5cm} \left. - \frac{\eta_0}{2} (x^2-1)
    \left(\frac{\arctan{\left(x\cot{\gamma_0}\right)}}{x\cot{\gamma_0}}
      - \frac{1}{1+x^2\cot^2{\gamma_0}}\right) \right\},
\end{split}
\end{equation}
where $G_{3D} = 32 \sqrt{\frac{\pi}{m}}~{a}^2~n_0~g_{3D}({2{a}})$ and
$g_{3D}({2{a}})\approx \frac{1-\nu/2}{(1-\nu)^3}$ \cite{carnahan69}.

\end{appendix} 

\end{document}